\documentclass[fleqn,usenatbib]{mnras}

\usepackage{newtxtext,newtxmath}

\usepackage[T1]{fontenc}

\DeclareRobustCommand{\VAN}[3]{#2}
\let\VANthebibliography\thebibliography
\def\thebibliography{\DeclareRobustCommand{\VAN}[3]{##3}\VANthebibliography}
\usepackage{xcolor}

\usepackage{graphicx}	
\usepackage{amsmath}	

\title[Dust evolution in self-gravitating discs]{Dust density enhancements and the direct formation of planetary cores in gravitationally unstable discs}

\author[Rice et al.]{
Ken Rice$^{1,2}$\thanks{E-mail: wkmr@roe.ac.uk}, Hans Baehr$^{3,4,5}$, Alison K. Young$^{6,7}$, Richard Booth$^{8}$, Sahl Rowther$^{9}$, 
\newauthor
Farzana Meru$^{6,7}$, Cassandra Hall$^{3,4}$, Adam Koval$^{1,2}$ \\
$^{1}$SUPA, Institute for Astronomy, University of Edinburgh, The Royal Observatory, Blackford Hill, Edinburgh, EH9 3HJ, UK \\
$^{2}$Centre for Exoplanet Science, University of Edinburgh, Edinburgh, EH9 3HJ, UK \\
$^{3}$Department of Physics and Astronomy, The University of Georgia, Athens, GA 30602, USA \\
$^{4}$Center for Simulational Physics, The University of Georgia, Athens, GA 30602, USA \\
$^{5}$Max Planck Institute for Astronomy, Königstuhl 17, D-69117 Heidelberg, Germany \\
$^{6}$ Centre for Exoplanets and Habitability, University of Warwick, Coventry CV4 7AL, UK \\
$^{7}$ Department of Physics, University of Warwick, Coventry CV4 7AL, UK \\
$^{8}$ School of Physics and Astronomy, University of Leeds, Leeds LS2 9JT, UK \\
$^{9}$ Department of Physics and Astronomy, University of Leicester, Leicester LE1 7RH, United Kingdom
}

\date{Accepted XXX. Received YYY; in original form ZZZ}

\pubyear{2015}

\begin{document}
\label{firstpage}
\pagerange{\pageref{firstpage}--\pageref{lastpage}}
\maketitle

\begin{abstract}
Planet formation via core accretion involves the growth of solids that can accumulate to form planetary cores.  There are a number of barriers to the collisional growth of solids in protostellar discs, one of which is the drift, or metre, barrier.  Solid particles experience a drag force that will tend to cause them to drift towards the central star in smooth, laminar discs, potentially removing particles before they grow large enough to decouple from the disc gas. Here we present 3-dimensional, shearing box simulations that explore the dynamical evolution of solids in a protostellar disc that is massive enough for the gravitational instability to manifest as spiral density waves.  We expand on earlier work by considering a range of particle sizes and find that the spirals can still enhance the local solid density by more than an order of magnitude, potentially aiding grain growth.  Furthermore, if solid particles have enough mass, and the particle size distribution extends to sufficiently large particle sizes, the solid component of the disc can undergo direct gravitational collapse to form bound clumps with masses typically between $1$ and $10$~M$_\oplus$. Thus, the concentration of dust in a self-gravitating disc could bypass the size barrier for collisional growth and directly form planetary cores early in the lifetime of the disc.
\end{abstract}




\section{Introduction}
Broadly speaking, there are two main planet formation scenarios, core accretion \citep{mizuno80} and direct collapse in gravitationally unstable discs \citep{kuiper51,boss97}.  Although the latter could form giant planets on wide orbits \citep{nero09,kratter10}, models and observations indicate that it probably rarely operates \citep{rice15,vigan17} and that most known planets likely form via core accretion \citep{schlaufman18}. 

The basic core accretion process involves the growth of initially micron-sized dust grains into much larger objects \citep{blum08} that can then coagulate to form large rocky bodies that either become terrestrial planets, or the cores of giant planets.  If the solid core becomes massive enough before the gas disc has dissipated, it can then undergo a phase of runaway gas accretion to form a giant planet \citep{pollack96}. 

There are, however, a number of barriers to the growth of solids in a protoplanetary disc, one of which is the drift, or metre, barrier \citep{laibe14}.  In a smooth, laminar protoplanetary disc, the gas will orbit with sub-Keplerian velocities because the negative pressure gradient slightly reduces the net inward radial force.  A consequence of this is that the solid particles, which are not influenced by the gas pressure, feel a head-wind that produces a drag force, causing these particles to lose angular momentum and drift inwards \citep{weidenschilling77}.  

The rate at which the solids drift inwards depends on their size.  Very small particles will be strongly coupled to the gas and will drift very slowly.  Very large particles will become decoupled from the gas and will also drift very slowly.  Intermediate-sized particles, however, can drift at speeds that are high enough for them to be lost to the central star before becoming large enough to no longer be strongly influenced by the drag force.  Although the exact size will depend on the disc properties, for typical protoplanetary discs, we'd expect the drift rate to peak for cm- to metre-sized particles \citep{weidenschilling77}.

The streaming instability \citep{youdin07,bai10} provides one possible way to overcome this drift-barrier. If regions with enhanced solid density start to develop, then the back-reaction of the particles on the gas can increase the velocity of the gas, resulting in a smaller velocity differential between the gas and the dust, therefore reducing the impact of gas drag, and slowing the local inward drift velocity of the particles. This region can then accumulate solids as faster drifting particles catch up with those whose drift has slowed, producing large enhancements in the local density of solids and potentially leading to gravitational collapse of the solid component of the disc, and the direct formation of planetesimals \citep{johansen09,simon16}.

However, it's also recognised that the presence of high-density structures, or pressure maxima, in the gas disc can also influence the inward drift of solid particles \citep{haghighipour03}. The drag force will cause particles to drift towards pressure maxima, so the presence of vortices \citep{godon99,johansen04,gibbons15} or even small-scale turbulence \citep{johansen06} can lead to the concentration of solid particles. 

During the earliest stages of protostellar evolution, protostellar discs may also be massive enough to be susceptible to the gravitational instability \citep{lin90,rice10}, which manifests as spiral density waves. Fragmentation into bound objects may also occur if the system is sufficiently unstable \citep[e.g.,][]{boss97,durisen07}.  However, there are indications from theoretical work and population synthesis models that fragmentation does not adequately explain the observed gas giant population \citep[e.g.,][]{rice15,forgan18,muller18}, suggesting that a self-gravitating phase is more likely to manifest as spiral density waves, which will also be sites where particles could concentrate.  Previous work \citep[e.g.,][]{rice04,gibbons12,booth2016,baehr21} has demonstrated that the local particle density can be substantially enhanced in self-gravitating spirals, potentially reaching densities where planetesimals could then form by direct gravitational collapse \citep{rice06,gibbons14,baehr22,longarini23b,rowther24a}.

However, most of this earlier work either assumed particles of a single size, or particles with a single stopping time. This is very useful for understanding how spirals will influence different types of particles, what types of particles would need to be present in order for the enhancements to be substantial, and to get some idea of the criteria required for the particles to undergo direct gravitational collapse. However, this is unrealistic, given that the particles typically have a size distribution that can cover many orders of magnitude. The outcome therefore depends on the range of particle sizes and also how much mass is in particles that are strongly influenced by the spiral density waves.   

For example, a grain size distribution dominated by smaller grains is expected to develop weaker concentrations within the spiral features \citep{gibbons14,shi16}, making the collapse into bound clumps more difficult. Furthermore, ever larger quantities of dust are required to overcome the dust diffusion for small dust grains since the collapse timescale of a cloud of dust increases for smaller grains due to their shorter stopping time and lower terminal velocity. Thus smaller dust grains are more likely to be prevented from collapsing by the same amount of diffusion \citep{klahr20}.

Here we use 3-dimensional shearing box simulations to investigate if spiral density waves can still substantially enhance the local dust density when a particle size distribution is adopted.  We start by considering a particle distribution that covers a wide range in particle size, but also narrow this down to focus on the regime where the particles are most strongly affected by the spirals. In particular, we investigate the conditions under which the particle component may undergo direct gravitational collapse to form dense clumps \citep[e.g.,][]{baehr22}.   

The paper is structured as follows.  In Section \ref{sec:methods} we introduce the {\sc pencil} code \footnote{\url{https://github.com/pencil-code}}, which is used for all the simulations presented here, and describe the simulation setup. Section \ref{sec:results} presents the results from the suite of simulations carried out, and analyses the output from these simulations.  Section \ref{sec:discussion} discusses these results and provides some broader context, and Section \ref{sec:conclusion} presents some concluding remarks.  

\section{Methods} \label{sec:methods}
The simulations presented here are 3-dimensional hydrodynamic shearing-box simulations using the {\sc pencil} code \citep{brandenburg03,pencilcodecoll}. The shearing-box simulations solve the continuity, momentum, and entropy equations in co-rotating Cartesian coordinates.  The centre of the box is assumed to be at some arbitrary radius from the central object and is rotating with the disc's angular frequency, $\Omega$, at this radius. The basic equations for gas density, $\rho_g$, gas velocity, $\bf{u}$, and entropy, $s$, are:
\begin{equation}
\frac{\partial \rho_{\rm g}}{\partial t} + \nabla \cdot (\rho_{\rm g} {\bf u}) - q \Omega x \frac{\partial \rho_{\rm g}}{\partial y} = f_D({\rho_{\rm g}})
\end{equation}
\begin{equation}
\frac{\partial {\bf u}}{\partial t} + {\bf u} \cdot (\nabla {\bf u}) - q \Omega x \frac{\partial {\bf u}}{\partial y} = - \frac{\nabla P}{\rho_{\rm g}} + q \Omega u_x \hat{y} - 2 \Omega \times {\bf u} - \nabla \Phi + f_\nu({\bf u})
\label{eq:gasvel}
\end{equation}
\begin{equation}
\frac{\partial s}{\partial t} + ({\bf u} \cdot \nabla) s - q \Omega x \frac{\partial s}{\partial y} = \frac{1}{\rho_{\rm g} T} \left(2 \rho_{\rm g} \nu {\bf S}^2 - \Lambda + f_\chi(s) \right).
\label{eq:entropy}
\end{equation}
In the above equations, $\Phi$, is the gravitational potential of the gas, $P$ is the gas pressure, $T$ is the gas temperature, and the final terms on the right-hand side of each equation are the hyperdissipation terms, which have the form
\begin{equation}
    f_x = \nu (\nabla^6 x),
\end{equation}
with constant $\nu = 2 H_{\rm g}^6 \Omega$ \citep{yang12}. 

As highlighted in \cite{baehr22}, the gravitational potential is solved in Fourier space by transforming the density to find the potential at wavenumber $k$ and then transforming back into real space.  The shear periodic boundary conditions are accounted for as in \cite{johansen07}. 

To take the shear velocity into account, the velocity has the form ${\bf u} = (v_x, v_y +q \Omega x, v_z)$, where $q = 1.5$ is the Keplerian rotation profile adopted here and $v_x, v_y, v_z$ are the components of the velocity relative to the shear flow. We use an equation of state with internal energy $\epsilon$ and specific heat ratio, $\gamma = c_p/c_v = 5/3$, such that the pressure, $P$, is given by
\begin{equation}
P = (\gamma - 1) \rho_{\rm g} \epsilon,
\end{equation}
where $\epsilon = c_v T$.  As Equation \ref{eq:entropy} indicates, the thermodynamic variable in the {\sc pencil} code is the entropy, $s$, but we can relate entropy, temperature and sound speed through
\begin{equation}
c_{\rm s}^2 = c_{\rm s,0}^2 \exp \left[\frac{\gamma s}{c_p} + (\gamma - 1) \ln \left(\frac{\rho_{\rm g}}{\rho_{\rm g,0}} \right) \right],
\end{equation}
where $c_{\rm s,0}$ is the initial uniform sound speed, and $\rho_{\rm g, 0}$ is the initial gas density.

The energy equation has a cooling term, $\Lambda$, that is implemented using the $\beta$-prescription \citep{gammie01}
\begin{equation}
\Lambda = \frac{\rho (c_{\rm s}^2 - c^2_{\rm s, irr})}{(\gamma - 1)t_{\rm c}},
\label{eq:cooling}
\end{equation}
where $t_{\rm c} = \beta \Omega^{-1}$ and $c^2_{\rm s, irr}$ is a term intended to represent some level of background irradiation.  A protostellar disc is susceptible to the gravitational instability if $Q = c_s \Omega/\pi G \Sigma_{\rm g} \sim 1$ \citep{toomre64}, where $\Sigma_{\rm g}$ is the surface density of the gas disc.  In the simulations presented here, we use $\beta = 10$ and set $c^2_{\rm s, irr}$ so that the Toomre $Q$ value settles to a value close to unity. This ensures that the disc will become gravitationally unstable, manifesting as spiral density waves \citep{lodato04,baehr21}, but won't undergo fragmentation \citep{gammie01, rice05}. 

The gas is vertically stratified and, as in \citet{baehr17}, the vertical gravitational acceleration has a sinusoidal profile that goes to 0 at the boundary.  This isn't entirely realistic, but has been found to improve stability. 

\subsection{Reference gas simulation}\label{sec:refgas}
All of the simulations presented here use a shearing box in which the initial sound speed is $c_{\rm s,0} = \pi$ and in which the angular velocity is $\Omega = 1$.  Hence, the initial disc scale height is $H_{\rm g} = c_{\rm s,0}/\Omega = \pi$.  The gas disc has an average surface density of $\Sigma_{\rm g} = 1$ and is initialised with a vertical Gaussian profile that matches the expected initial gas scaleheight of $H_g = \pi$.  Using the approximation that $\Sigma_{\rm g} \sim \rho_{\rm g} 2 H_{\rm g}$ gives a density in the midplane of $\rho_{\rm g} \sim 0.16$.  The irradiation term in the cooling formalism (see Equation \ref{eq:cooling}) is set to $c_{\rm s, irr} = c_{\rm s,0} = \pi$ so that the disc will tend to cool to a state where $Q \sim 1$.

The box itself has a size of $L_x = 120$, $L_y = 120$, and $L_z = 20$ and a resolution of $256 \times 256 \times 128$.  This means that in the midplane ($x,y$) the box covers a region of about $40$ scale heights, while in the vertical direction, the box covers about $4-5$ scale heights, or just over $2$ scale heights above and below the midplane.  This does mean that the simulations do not cover a large number of vertical scale heights, but this was chosen to provide reasonable vertical resolution in the midplane. 

\begin{figure}
    \centering
    \includegraphics[width=0.49\textwidth]{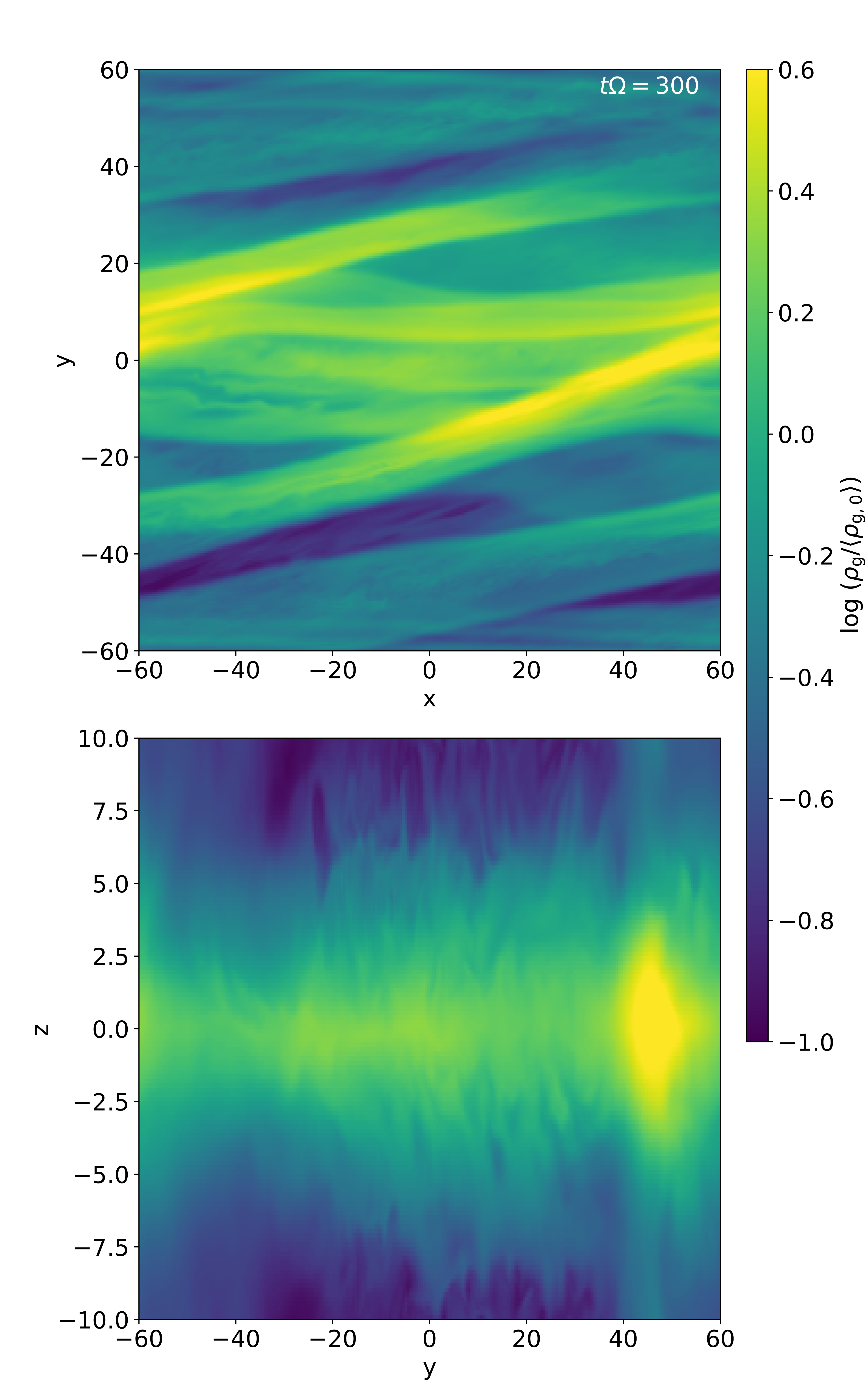}
    \caption{Gas density structure in the midplane (top panel) and a vertical slice through the disc at $x = 0$ (bottom panel) at $t \Omega = 300$.  The disc has settled into a quasi-steady, self-gravitating state with spiral density waves and the simulation is clearly able to capture the vertical structure in the gas disc.}
    \label{fig:refgas}
\end{figure}

Figure \ref{fig:refgas} shows the gas density structure in the midplane (top panel) and a vertical slice through the disc at $x = 0$ (bottom panel) at a time of $t \Omega = 300$.  The disc has settled into a quasi-steady, self-gravitating state with spiral density waves.  Even though the simulations only cover a few scale heights in the vertical direction, the vertical structure of the gas disc is still reasonably well captured. 

\subsection{Including dust}
The {\sc pencil} code allows for the inclusion of dust, or solid particles. The solid particles are Lagrangian superparticles embedded in the Eulerian mesh \citep{brandenburg03}. The particles interact with the gas via a drag force and via gravity. The backreaction of the particles on the gas can also be included. Apart from the test particle simulation, all the simulations presented here include the backreaction of the solid particles on the gas. 

Since the particles do not feel the gas pressure, the equation of motion for particle $i$ is \citep{yang16}:
\begin{equation}
\frac{d {\bf w}_i}{dt}=2 \Omega w_{y,i} {\bf \hat{x}} + (q - 2) \Omega w_{x,i} {\bf \hat{y}} - \nabla\Phi + \frac{1}{\tau_s} \left({\bf w}_i - {\bf u}({\bf x}_i) \right) 
\end{equation}
\begin{equation}
\frac{d {\bf x}_i}{dt} = {\bf w}_i - q \Omega x_i {\bf \hat{y}},
\end{equation}
where ${\bf w}_i$ is the velocity of particle $i$, ${\bf x}_i$ is the position of particle $i$, and $\tau_s$ is the drag force stopping time. The gas velocity at the location of particle $i$, ${\bf u}({\bf x}_i)$, is determined using a second-order spline interpolation often referred to as the Triangular Shaped Cloud scheme \citep{youdin07}.

In the Epstein regime, the stopping time is \citep{weidenschilling77}
\begin{equation}
\tau_{\rm s} = \frac{a \rho_{\rm d}}{c_s \rho_{\rm g}},
\label{eq:stoppingtime}
\end{equation}
where $a$ is the particle diameter, and $\rho_{\rm d}$ is the physical density of the dust particles, which is taken to be a constant.  If we normalise by the dynamical time, $\Omega^{-1}$, we get the Stokes number,
\begin{equation}
\mathrm{St} = \tau_{\rm s} \Omega.
\label{eq:Stokes}
\end{equation}
Since $\Omega = 1$ in these simulations, the stopping time and Stokes numbers are the same.  

In these simulations, the particles are superparticles, a collection - or swarm - of identical dust particles that do not move with respect to each other.  The superparticles can be setup to have pre-specified stopping times, or can be initialised with pre-specified sizes, $a$, which is what we consider here.  There are a number of different possible initial size distributions, but in this work we assume a power-law size distribution with a power-law index that gives an equal number of superparticles in each logarithmic size bin.  This implies a grain size distribution of $n(a) \propto a^{-4}$ which is slightly steeper than the canonical interstellar medium (ISM) grain size distribution of $n(a) \propto a^{-3.5}$ \citep{mathis77}, meaning a larger fraction of the mass is in the smaller grains. In contrast, a grain size distribution shallower than $n(a) \propto a^{-3.5}$ would mean more mass is in the larger particles, implying some amount of grain growth. 

The reason for choosing this size distribution is partly because it is close to that expected and partly computational.  Since each superparticle has the same mass, an equal mass in each logarithmic size bin also means the same number of particles in each logarithmic size bin.  A shallower size distribution would reduce the resolution in the smallest size bins, while a steeper one would reduce the resolution in the largest size bins.   
 
The particles can be assumed either to be test particles, or to have a total mass that is some fraction of the total gas mass.  The particle self-gravity can also be included, as can the back-reaction of the particles on the gas.  When including the back-reaction of the solid particles on the gas, a drag force of the form $-({\bf w}_i - {\bf u}({\bf x}_i))/\tau_s$ is included on the right-hand side of Equation \ref{eq:gasvel}.  When including the particle self-gravity, the density of the gas and dust are combined to produce a potential $\Phi = \Phi_{\rm g} + \Phi_{\rm d}$, so that the self-gravity of both the gas and particles is included and the gravitational potential influences both components.  

We consider a scenario in which the particles are treated as test particles, and also scenarios where both the particle self-gravity and the backreaction of the particles on the gas is included.  

\section{Results} \label{sec:results}
We consider a number of different simulations, starting with one in which the solid particles are taken to be test particles whose self-gravity is not included and in which there is no back-reaction on the gas.  We then carry out a suite of simulations in which the particle self-gravity and back-reaction on the gas are included, varying the total mass of the particles - relative to the gas - from $10^{-4}$ to $0.025$. Each simulation has $10^7$ superparticles, each of which has the same mass, and, as mentioned above, the size distribution is chosen so that there is an equal mass in each logarithmic size bin. The superparticles are initially distributed with a Gaussian vertical density profile that is close to that of the initial vertical profile of the gas disc.   

We consider a range of particles sizes, which we represent as approximately the Stokes number that the particle would have if the density and sound speed were the average values in the midplane once the gas disc has settled into a quasi-steady state (e.g., Equation \ref{eq:Stokes}).  We should stress, though, that these are representative Stokes numbers and that the particles' actual Stokes numbers will depend on the local gas density and sound speed. Table \ref{tab:sims} presents the basic parameters of all the simulations considered here.

\begin{center}
\begin{table}
\centering
    \caption{Table showing the simulations presented here. The term $m_{\rm d}/m_{\rm g}$ is the total mass of the solid particles relative to the total gas mass, and $\mathrm{St_{min}}$ and $\mathrm{St_{max}}$ are the lower and upper limits to the representative Stokes numbers of the particles considered. The term {\em clumps} indicates whether or not dense clumps formed in that simulation. }
\begin{tabular}{|l|c|c|c|c|} 
 \hline
 Simulation & $m_{\rm d}/m_{\rm g}$ & $\mathrm{St_{min}}$ & $\mathrm{St_{max}}$ & clumps\\ 
 \hline\hline
 md0St002-200 & 0 & 0.02 & 200 & No\\ 
 \hline
 md00001St002-200 & 0.0001 & 0.02 & 200 & No \\
 \hline
 md0001St002-200 & 0.001 & 0.02 & 200 & No\\
 \hline
 md003St002-200 & 0.003 & 0.02& 200 & No \\
 \hline
 md007St002-200 & 0.007 & 0.02 & 200 & No \\
 \hline
 md01St002-200 & 0.01 & $0.02$ & $200$ & Yes \\
 \hline
 md025St002-200 & 0.025 & $0.02$ & $200$ & Yes\\
 \hline
 md001St01-1 & 0.001 & $0.1$ & $1$ & No \\
 \hline
 md003St01-1 & 0.003 & $0.1$ & $1$ & No \\
\hline
 md007St01-1 & 0.007 & $0.1$ & $1$ & Yes \\
 \hline
 md001St1-10 & 0.001 & $1$ & $10$ & No \\
 \hline
 md003St1-10 & 0.003 & $1$ & $10$ & Yes \\
\hline
 md007St1-10 & 0.007 & $1$ & $10$ & Yes \\
 \hline
 md01St002-02 & 0.01 & $0.02$ & $0.2$ & No \\
 \hline
 md01St01-1 & 0.01 & $0.1$ & $1$ & Yes \\
 \hline
 md01St1-10 & 0.01 & $1$ & $10$ & Yes \\
 \hline
 md01St2-20 & 0.01 & $2$ & $20$ & Yes \\
 \hline
 md01St10-100 & 0.01 & $10$ & $100$ & No \\
 \hline
 md01St20-200 & 0.01 & $20$ & $200$ & No \\
 \hline
\end{tabular}
\label{tab:sims}
\end{table}
\end{center}

\subsection{Test particle simulation} \label{sec:refdust}
In this initial simulation, the particles are taken to be test particles and, as mentioned above, this means that their self-gravity and the back-reaction of the particles on the gas is not included. However, the particles do feel the gravity of the gas and do experience a drag force, which we assume is in the Epstein regime.  However, the simulations start with no gas drag, which is then turned on at $t \Omega = 40$.  This ensures that the gas disc has reached a quasi-steady state before we introduce the drag force that will act to concentrate the solids in the resulting density structures.

For this test particle simulation, the size distribution varies from $a = 0.01/ \rho_{\rm d}$ to $a = 100/ \rho_{\rm d}$.  In the reference gas simulation (see Section \ref{sec:refgas} and Figure \ref{fig:refgas}) the mean midplane gas density is $\rho_{\rm g} \sim 0.1$ and the mean sound speed, once a quasi-steady state has been reached, is $c_{\rm s} \sim 5.0$.  Equations \ref{eq:stoppingtime} and \ref{eq:Stokes} indicate that the stopping time, and Stokes number, will also vary over 4 orders of magnitude, from about $\mathrm{St} \sim 0.02$ to $\mathrm{St} \sim 200$.  However, the local stopping time/Stokes number will depend on the local gas density and sound speed, which can differ from these mean values.  Hence, the Stokes numbers that we quote will be a representative approximation of the Stokes numbers for the particles being considered, but will not be their exact Stokes numbers.  

Equations \ref{eq:stoppingtime} and \ref{eq:Stokes} show that the Stokes number of particles of a given size, $a$, depends on the local gas density, $\rho_{\rm g}$, and local gas sound speed, $c_s$.  Since the sound speed tends to be higher in the spiral arms than in the inter-arm regions, the Stokes number for a given particle size can vary quite substantially. Once the gas disc has settled into a quasi-steady state, the Stokes number of the particles we class as $\mathrm{St} \sim 1$ can vary from $\mathrm{St} = 0.1$ to $\mathrm{St} = 12$, but most (95\%) lie in the range $\mathrm{St} \sim 0.3 - 5$. 

\begin{figure}
    \centering
    \includegraphics[width=0.495\textwidth]{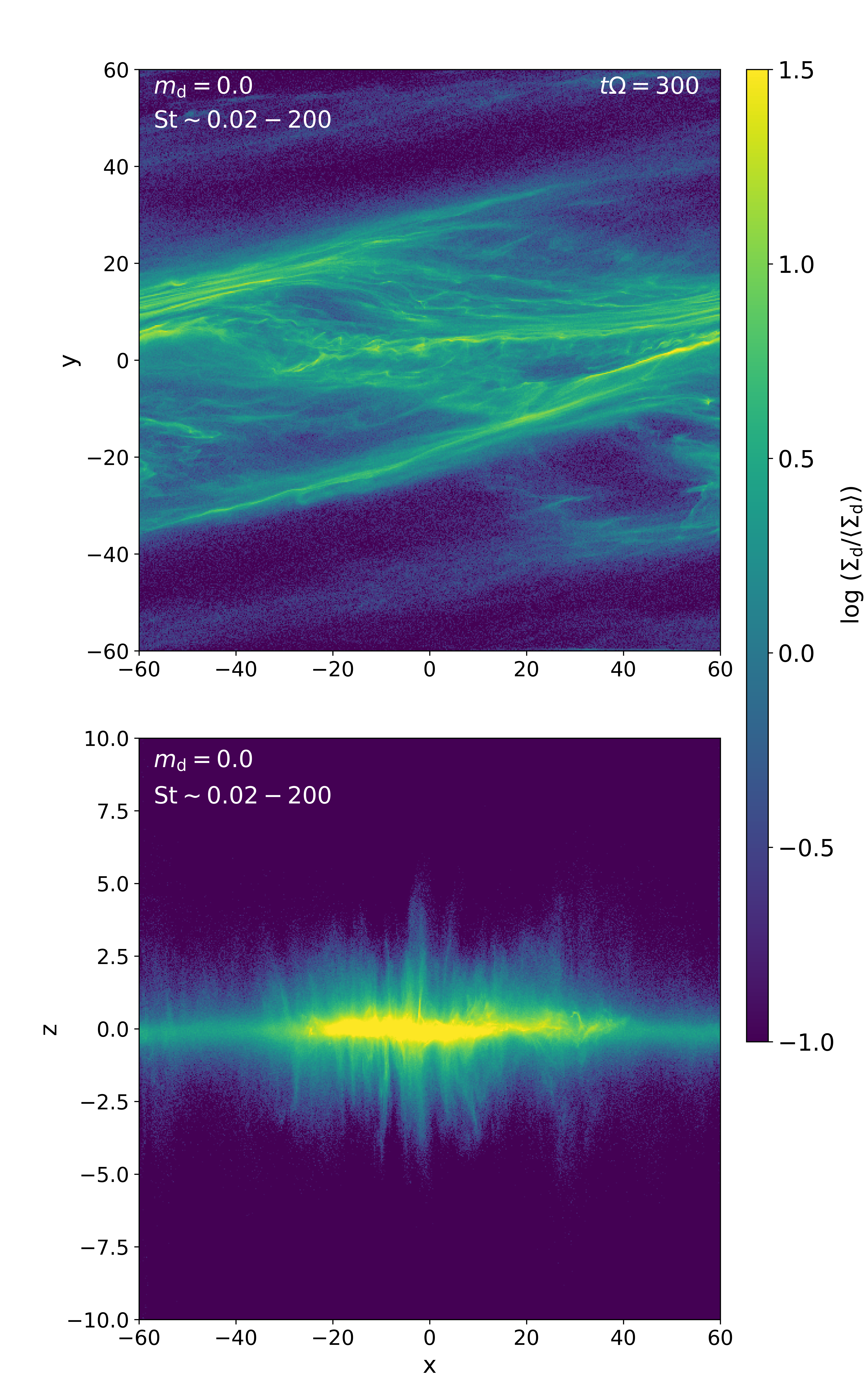}
    \caption{Surface density structure of the particle disc at $t \Omega = 300$, for the full size range of solid particles from the test particle simulation.  The top panel shows the particle surface density in the disc midplane, while the bottom panel shows the surface, or column, density projected on to the xz-plane.}
    \label{fig:refdust}
\end{figure}

Figure \ref{fig:refdust} shows the surface density of the solid particles at the same simulation time as shown in Figure \ref{fig:refgas}.  The top panel shows the surface density in the xy-plane, while the bottom panel is the column density projected onto the xz-plane.  We note that for the solid particles we present surface, or column, densities, rather than volume densities.  Figure \ref{fig:refdust} illustrates that the spiral features in the dust can be much narrower than in the gas (c.f., Figure \ref{fig:refgas}), and that the dust will tend to settle towards the midplane, resulting in a smaller scale height than that of the gas.    

\begin{figure}
    \centering
    \includegraphics[width=0.495\textwidth]{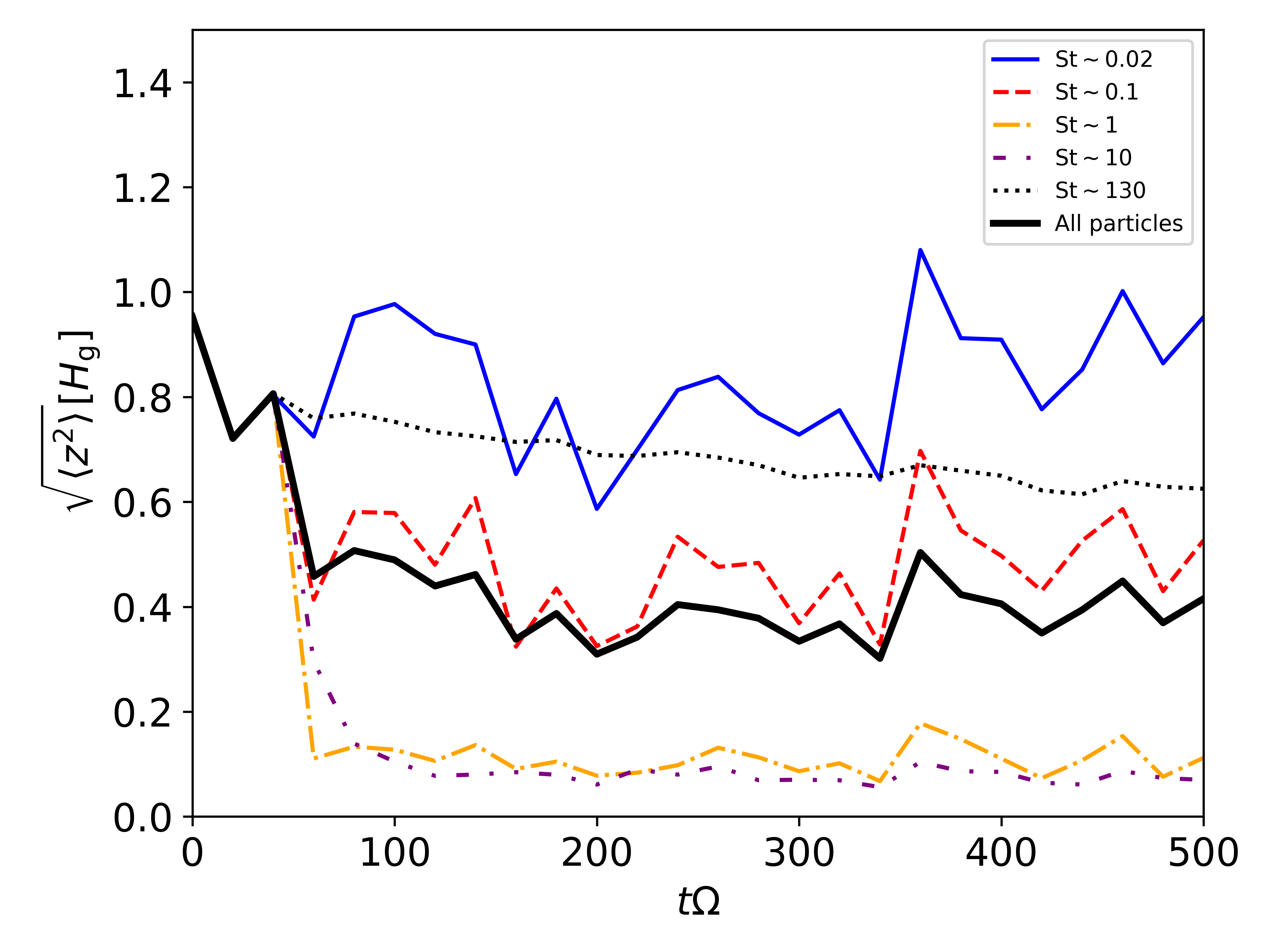}
    \caption{Scaleheight of the solid particles, as represented by the root mean square of the particles' vertical positions normalised by the initial scaleheight of the gas disc ($H_{\rm g} = \pi$).  The thick solid line shows all the particles, while the other lines shows particles with Stokes numbers of $\mathrm{St} \sim 0.02$ (thin solid line), $\mathrm{St} \sim 0.1$ (dashed line), $\mathrm{St} \sim 1$ (dash-dot line), $\mathrm{St} \sim 10$ (dash-dot-dot line), and $\mathrm{St} \sim 130$ (dotted line). The particles all start with a scaleheight similar to the initial scaleheight of the gas disc.  The smallest ($\mathrm{St} \sim 0.02$) and largest ($\mathrm{St} \sim 130$) particles largely retain this initial scaleheight, while the $\mathrm{St} \sim 1$ and $\mathrm{St} \sim 10$ particles settle to a very thin layer in the disc midplane.}  
    \label{fig:dustscaleheight}
\end{figure}

Figure \ref{fig:dustscaleheight} shows how the scaleheight of the solids varies with particle size, or Stokes number.  The figure shows the root mean square of the vertical position of the dust particles, scaled by the initial scaleheight, $H_{\rm g} = \pi$, of the gas disc.  Figure \ref{fig:dustscaleheight} shows all the particles (thick solid line), and then selects narrow ranges of particle size to select particles with Stokes numbers of $\mathrm{St} \sim 0.02$ (thin solid line), $\mathrm{St} \sim 0.1$ (dashed line), $\mathrm{St} \sim 1$ (dash-dot line), $\mathrm{St} \sim 10$ (dash-dot-dot line), and $\mathrm{St} \sim 130$ (dotted line). The particle size range is selected to cover a factor of 1.5 in particle size/Stokes number. Again, the Stokes numbers are representative and the actual values would depend on the local gas density and sound speed.

Figure \ref{fig:dustscaleheight} shows that the particles all start with a scaleheight very close to the initial scaleheight in the gas disc.  There is initially no gas drag and it is turned on at $t \Omega = 40$. The smallest particles (thin solid blue line) retain this initial scaleheight and don't undergo any settling. The largest with Stokes numbers of $\mathrm{St} \sim 130$ (dotted black line) remain close to this initial scaleheight but are slowly settling as their vertical motion is slowly damped by the drag force. Particles, with Stokes numbers of $\sim 0.1$ (dashed red line) do undergo some settling,  while intermediate-sized particles, with Stokes numbers between about $\mathrm{St} \sim 1$ (dash-dot yellow line) and $\mathrm{St} \sim 10$ (dash-dot-dot purple line), settle into a very thin layer around the midplane.

When considering all the particles together (thick solid line), they settle into a layer with a scaleheight about half their original scaleheight, or about one-quarter of the gas scaleheight once the disc has settled into a quasi-steady state. This is also illustrated by the bottom panel in Figure \ref{fig:refdust}.  This will, however, depend on the range of Stokes numbers considered \citep[e.g.,][]{youdinlithwick07}.  If we were considering a much narrower range of Stokes numbers around unity, we'd expect the particle disc to be much thinner.  Conversely, if we were considering a much wider range of Stokes numbers, we'd expect the scaleheight of the particle disc to be larger.  

\begin{figure*}
    \centering
    \includegraphics[width=0.95\textwidth]{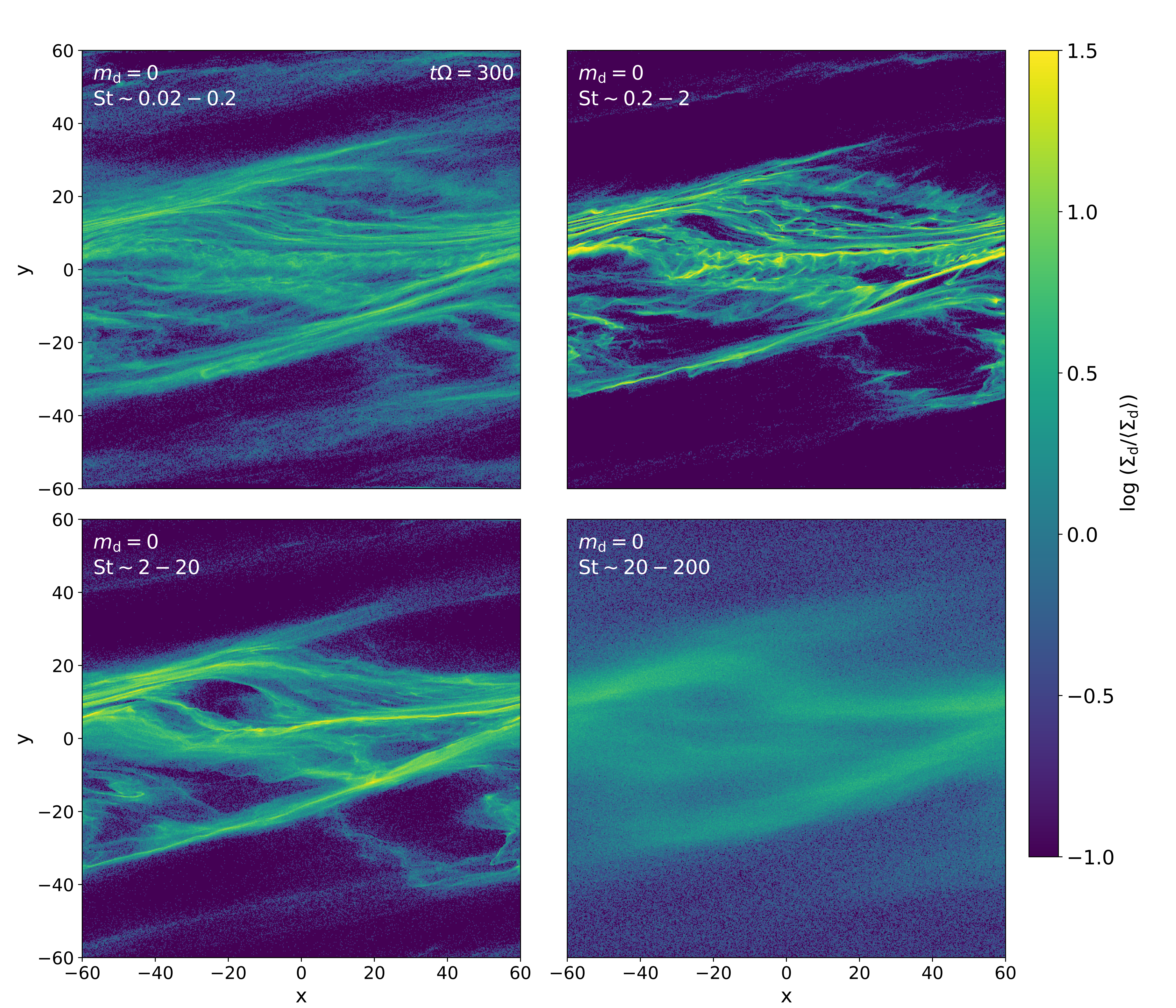}
    \caption{Surface density structure of the solid particles in the test particle simulation at $t \Omega = 300$ and for particles of different sizes. Each panel considers one order of magnitude in particle size, which are presented as approximate Stokes number ranges. The smallest are in the top-left panel, and the largest in the bottom-right. The smallest largely trace the structures in the gas disc (see Figure \ref{fig:refgas}), while the largest show some of the gas structure, but are becoming decoupled from the disc gas.  For particles with Stokes numbers around unity (top-right and bottom-left panels) the structures in the particles are much narrower than the corresponding structures in the gas and the local particle surface density can be significantly enhanced.}
    \label{fig:dust_Stokes}
\end{figure*}

The surface density structure shown in Figure \ref{fig:refdust} is for all of the particles in the simulation.  However, as already illustrated in Figure \ref{fig:dustscaleheight}, particles of different sizes, or Stokes numbers, will evolve differently. In Figure \ref{fig:dust_Stokes}, we consider 4 different particle size ranges.  In each case, the surface density has been normalised by the average surface density for the particles being considered.  The top-left panel is particles with $a/\rho_d = 0.01 - 0.1$, the top-right panel is $a/\rho_d = 0.1 - 1$, the bottom left panel is $a/\rho_d = 1 - 10$, and the bottom-right panel is $a/\rho_d = 10 - 100$, corresponding to representative Stokes number ranges of $\mathrm{St} \sim 0.02 - 0.2$, $\mathrm{St} \sim 0.2 - 2$, $\mathrm{St} \sim 2 - 20$, and $\mathrm{St} \sim 20 - 200$, respectively.  

Figure \ref{fig:dust_Stokes} illustrates that the particles with small Stokes numbers (top-left panel) largely trace the structures in the gas disc (see Figure \ref{fig:refgas}). As you approach Stokes numbers of order unity (top-right panel) and for Stokes slightly larger than unity (bottom-left panel), the features in the particle disc can become much narrower than the corresponding feature in the gas disc, and the local particle surface density can become significantly enhanced. This is partly because, as shown in Figure \ref{fig:dustscaleheight}, these particles will undergo more settling than the smaller, and larger, particles, and partly because the drag force will also tend to concentrate them in pressure maxima (e.g., \citealt{rice04}). At larger Stokes numbers (bottom-right panel), the particles start to decouple from the gas, and although they do trace some of the features in the gas disc, the enhancement in the local density of solids is relatively small.  

\subsubsection{Surface density enhancements}
\begin{figure}
    \centering
    \includegraphics[width=0.48\textwidth]{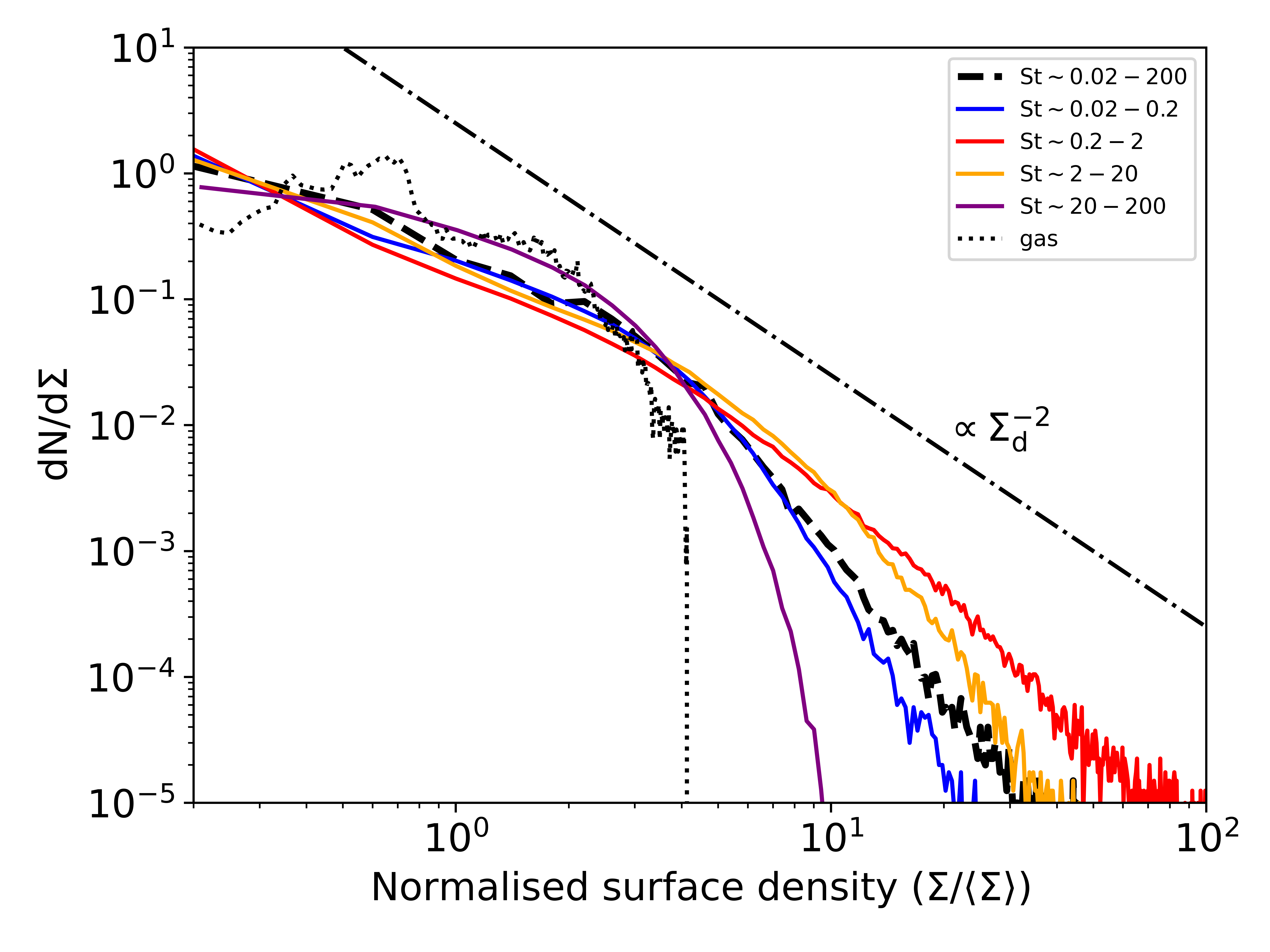}
    \caption{Distribution of surface densities for the gas (dotted line) and solid particles (solid lines and dashed line) in the test particle simulation at $t \Omega = 300$.  The gas surface density can vary by a factor of a few relative to the average.  When considering all of the particles (black dashed line) there can be regions where the surface density is enhanced by more than an order of magnitude.  For particles with Stokes numbers near unity (red and orange lines) this enhancement is more substantial, potentially reaching values two orders of magnitude greater than the average surface density of these particles. For the smallest and largest particles (blue and purple lines) the enhancement is smaller.}
    \label{fig:sdensdistr}
\end{figure}

Figure \ref{fig:sdensdistr} shows the distribution of surface densities in the gas (dotted line) and in the solid particles (solid lines and dashed line) in the test particle simulation at $t \Omega = 300$.  For the gas, we simply use the existing grid and integrate the gas density along the vertical column to get the gas surface density in the $x-y$ plane.  The values in each grid cell are then normalised by the average gas surface density, $\left< \Sigma_g \right> = 1$.  Figure \ref{fig:sdensdistr} shows that the local gas surface density can vary by a factor of a few relative to the mean gas surface density. 

For the solid particles, we construct a $1000 \times 1000$ grid in the $x-y$ plane.  The surface density in each grid cell is simply the sum of the particles that would lie in that grid cell if they were located at $z = 0$. The surface density is then normalised by the mean surface density of the particles being considered.  Figure \ref{fig:sdensdistr} shows the resulting distribution of surface densities in this two-dimensional grid, for all the particles in the simulation (black dashed line) and for $4$ different ranges of particle sizes, each of which span an order of magnitude in particle size and which are presented as representative Stokes number ranges.

When considering all the particles (black dashed line) the local surface density can be enhanced by up to a factor of $\sim 30$.  For the smallest particles ($\mathrm{St} \sim 0.02 - 0.2$) and largest particles ($\mathrm{St} \sim 20 - 200)$ the enhancement is generally smaller.  For the smallest, this is because the particles are more strongly coupled to the gas, while for the largest it is because they are becoming decoupled from the gas.  For particles with Stokes numbers around unity, the enhancement can be much more substantial, potentially reaching values two orders of magnitude greater than the average surface density of these particles.  Particles in this intermediate size range will be the most strongly perturbed by turbulent gas motions leading to the most efficient concentration in pressure maxima.

What Figure \ref{fig:sdensdistr} suggests is that even if the particles cover a wide range of Stokes numbers, there can be regions where the local particle surface density is enhanced by an order of magnitude, or more.  Of course, this will depend on the overall range of particle sizes and the size distribution.  

To provide some context, if the Stokes number $1$ particles have sizes of order a cm, then our broad size distribution extends from about $0.1$~mm to about $1$~m.  In a realistic disc, we might expect the size distribution to extend down to about $1$~$\mu$m, implying that we are ignoring about 2 orders of magnitude in particle size. This could suggest that we might be over-estimating the overall enhancements in particle surface densities, since the smallest particles will be strongly coupled to the gas and show enhancements similar to that of the gas. 

However, even for the $n(a) \propto a^{-4}$ size distribution considered here, most of the mass will be in the size range considered and the impact of not directly simulating the smallest particles should be relatively small.  Hence, even if we were able to consider a full particle size range (i.e., from $\sim 1 \mu{\rm m}$ to $\sim 1{\rm m}$) we'd still expect that there would be regions where the particle surface density was enhanced by at least a factor of $10$.

\subsection{Massive particles}
The test particle simulation discussed in Section \ref{sec:refdust} illustrates that if the solid particle size distribution covers a wide range of Stokes numbers, the local surface density can be enhanced by just over an order of magnitude, and that this can increase if the particles become more concentrated around a Stokes number of 1. We consider here if this is different if the particles have mass, their self-gravity is included, and if the back-reaction of the solid particles on the gas is also included.  

We start by repeating the simulation above, but now include the particle self-gravity and the back-reaction of the solid particles on the disc gas. We consider total particle masses that are $10^{-4}$, $10^{-3}$, $0.003$, $0.007$, $0.01$, and $0.025$ that of the gas. Again, we initially consider a size distribution that covers $4$ orders of magnitude in particle size and that has an equal mass in each logarithmic size bin.  Although the Stokes number of each particle will depend on the local gas density and sound speed (see Equations \ref{eq:stoppingtime} and \ref{eq:Stokes}), the chosen size distribution will again cover representative Stokes numbers that vary from $\mathrm{St} \sim 0.02$ to $\mathrm{St} \sim 200$.  In these simulations, as with those in Section \ref{sec:refdust}, the drag force is turned on at $t \Omega = 40$, again to allow the gas disc to reach a quasi-steady state before dust particles begin to concentrate.

\begin{figure*}
    \centering
    \includegraphics[width=0.95\textwidth]{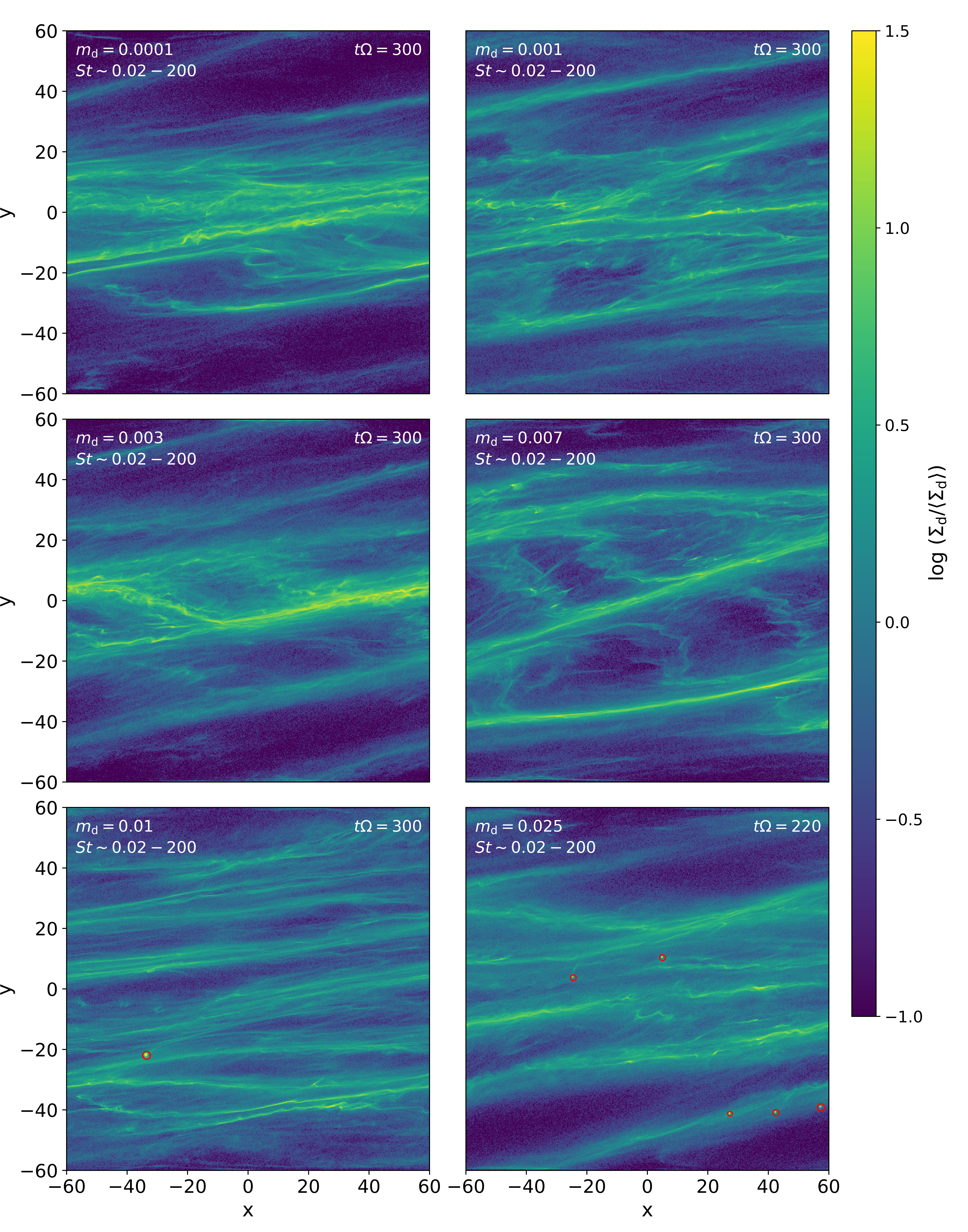}
    \caption{Surface density structure for all the particles in the simulation in which the particles have a mass $10^{-4}$ (top left), $10^{-3}$ (top right), $0.003$ (middle left), $0.007$ (middle right), $0.01$ (bottom left) and $0.025$ (bottom right) that of the gas. The particle self-gravity, and the backreaction on the gas, is included, and the particle size distribution covers 4 orders of magnitude in particle size with representative Stokes numbers ranging from $\mathrm{St} \sim 0.02$ to $\mathrm{St} \sim 200$. Apart from the $m_{\rm d} = 0.025 m_{\rm g}$ simulation, these are all at $t\Omega = 300$. For $m_{\rm d} = 0.025 m_{\rm g}$ the high densities of the clumps reduced the timestep and the simulation was stopped at $t \Omega = 220$. The red circles in the bottom two panels highlight dense clumps that have formed in these simulations.}
    \label{fig:ptclsfullrange}
\end{figure*}

Figure \ref{fig:ptclsfullrange} shows the surface density structure for all of the particles in the simulations with total particle masses of $m_{\rm d} = 10^{-4} m_{\rm g}$ (top left), $m_{\rm d} = 10^{-3} m_{\rm g}$ (top right), $m_{\rm d} = 0.003 m_{\rm g}$ (middle left), $m_{\rm d} = 0.007 m_{\rm g}$ (middle right), $m_{\rm d} = 0.01 m_{\rm g}$ (bottom left), and $m_{\rm d} = 0.025 m_{\rm g}$ (bottom right), where $m_{\rm g}$ is the total gas mass.  In each case, the particle surface density is normalised by the average particle surface density in that simulation.   

For the lower particle masses, the results are very similar to that for the test particle simulation shown in Figure \ref{fig:refdust}.  The local particle surface density can be enhanced by more than an order of magnitude, but there are no indications of dense clumps starting to form.  However, for $m_{\rm d} = 0.01 m_{\rm g}$ there is a single dense clump, highlighted by the red circle, and for $m_{\rm d} = 0.025 m_{\rm g}$ there are a number of dense particle clumps, also illustrated by red circles.  As discussed in more detail in Section \ref{sec:clumpid}, these clumps are identified as regions where the local particle surface density is at least 10 times the average surface density of the gas disc. The size of the red circles highlighting the clumps is then set by the Hill radius of the clump, $R_{\rm Hill} = (m_{\rm cl} R^3/3 M_*)^{1/3}$ where $m_{\rm cl}$ is the mass of the clump and $R$ is the assumed orbital radius of the centre of the shearing box. Since $\Omega = \sqrt{G M_*/R^3} = 1$ and $G = 1$, the Hill radius in code units becomes $R_{\rm Hill} = (m_{\rm cl}/3)^{1/3}$.

That clumps have formed in the $m_{\rm d} = 0.01 m_{\rm g}$ and $m_{\rm d} = 0.025 m_{\rm g}$ simulations demonstrates that even if we consider a wide range of particle sizes, if the dust-to-gas ratio is of order $0.01$, the canonical dust-to-gas ratio of the interstellar medium, there can be regions where the particle density becomes high enough for gravitational collapse to directly form dense clumps.  

\subsubsection{Narrowing the size range}
Figure \ref{fig:ptclsfullrange} is broadly consistent with earlier work \citep{gibbons14,baehr22,longarini23a, longarini23b,rowther24a} suggesting that to collapse to form dense clumps, the dust-to-gas ratio in the disc needs to be of order $0.01$ and the particles must have Stokes numbers near unity.  Here, however, the particles have a wide range of Stokes numbers ($\mathrm{St} \sim 0.02 - 200$) unlike many of the earlier studies in which the particles have either had a fixed size \citep[e.g.,][]{rowther24a} or a fixed Stokes number \citep[e.g.,][]{gibbons14,baehr22}.

To further explore the conditions under which the particles may start to form dense clumps, we consider reduced particle size ranges. Specifically, we reduce the size range so that it covers $\mathrm{St} \sim 0.1 - 1$ and $\mathrm{St} \sim 1 - 10$, rather than the $\mathrm{St} \sim 0.02 - 200$ considered above. This is motivated by Figure \ref{fig:sdensdistr} suggesting that for a size range that spans one order of magnitude, these are the particles that would probably experience the greatest enhancements in local surface density.

\begin{figure*}
    \centering
    \includegraphics[width=0.95\textwidth]{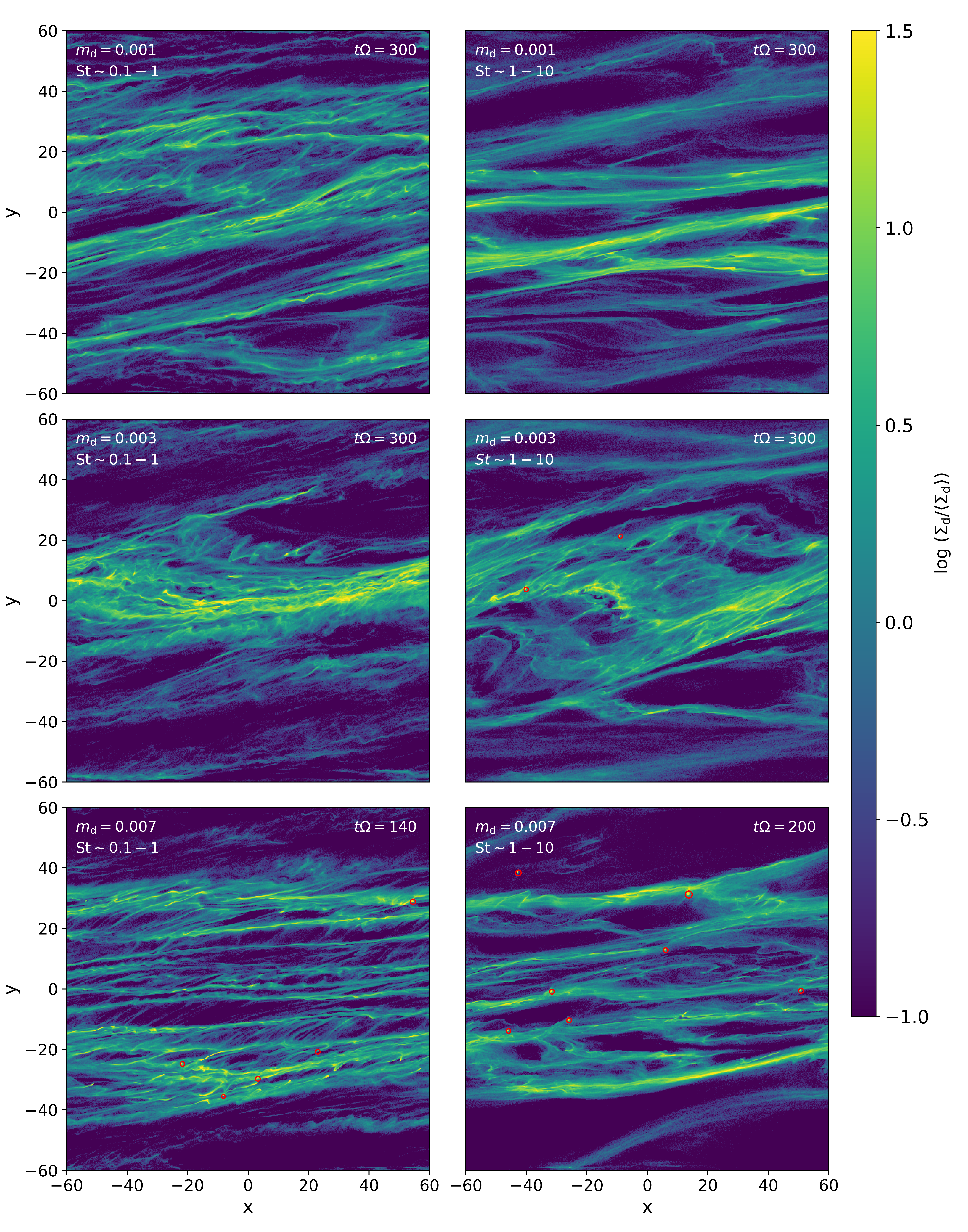}
\caption{Surface density structure of the solid particles in the simulations in which the Stokes number ranges are $\mathrm{St} \sim 0.1 - 1$ (left-hand panels) and $\mathrm{St} \sim 1 - 10$ (right-hand panels) and the total particles mass - from top to bottom - is $0.001$, $0.003$, and $0.007$ that of the gas. In each panel, the solid surface densities are normalised to the mean solid particle surface density in that simulations. The red circles highlight dense particle clumps that have formed in the discs.}
    \label{fig:n_5E14-5E13}
\end{figure*}

Figure \ref{fig:n_5E14-5E13} shows the surface density structures of the particles in simulations in which the Stokes number ranges from $\mathrm{St} \sim 0.1 - 1$ (left panels) and $\mathrm{St} \sim 1 - 10$ (right panels) and in which the total particles mass is $0.001$, $0.003$, and $0.007$ (top to bottom) that of the gas.  The simulations in which there are no dense clumps are all run to $t \Omega = 300$, as is the $m_{\rm d} = 0.003 m_{\rm g}$, $\mathrm{St} \sim 1 - 10$ simulation.  The dense clumps in the $m_{\rm d} = 0.007 m_{\rm g}$ simulations substantially reduce the timesteps and these simulations are stopped at $t \Omega = 140$ ($\mathrm{St} \sim 0.1 - 1$) and $t \Omega = 200$ ($\mathrm{St} \sim 1 - 10$).    

Although there is some small-scale structure in the $m_{\rm d} = 0.001 m_{\rm g}$ simulations, no dense clumps emerge.  Interestingly, when increasing the dust mass to $m_{\rm d} = 0.003 m_{\rm g}$, dense clumps start to emerge in the $\mathrm{St} \sim 1 - 10$ simulation, and for $m_{\rm d} \ge 0.007 m_{\rm g}$ we see clumps for both $\mathrm{St} \sim 0.1 - 1$ and $\mathrm{St} \sim 1 - 10$.
As in Figure \ref{fig:ptclsfullrange}, these clumps are highlighted by red circles that have a size set by the Hill radius of each clump.

This is broadly consistent with the results shown in Figure \ref{fig:ptclsfullrange}. When the range of particle sizes covers 4 orders of magnitude and covers Stokes numbers from $\mathrm{St} \sim 0.02 - 200$, dense clumps only form when the total particle mass is at least $1\%$ of the gas.  These simulations have the same particle mass in each logarithmic size bin.  Hence, in the simulation with $m_{\rm d} = 0.01 m_{\rm g}$ and $\mathrm{St} \sim 0.02 -200$, the particles with Stokes numbers in the range $\mathrm{St} \sim 0.1 - 1$ and in the range $\mathrm{St} \sim 1 - 10$ each have a total mass $m_{\rm d} = 0.0025 m_{\rm g}$.  This suggests that for the solid component of the disc to fragment to form dense clumps, there must be (a) particles with $\mathrm{St} \sim 1$ and (b) the dust-to-gas mass ratio for particles $0.5\lesssim \rm{St} \lesssim 5$ needs to be just over $10^{-3}$.

\begin{figure*}
    \centering
    \includegraphics[width=0.95\textwidth]{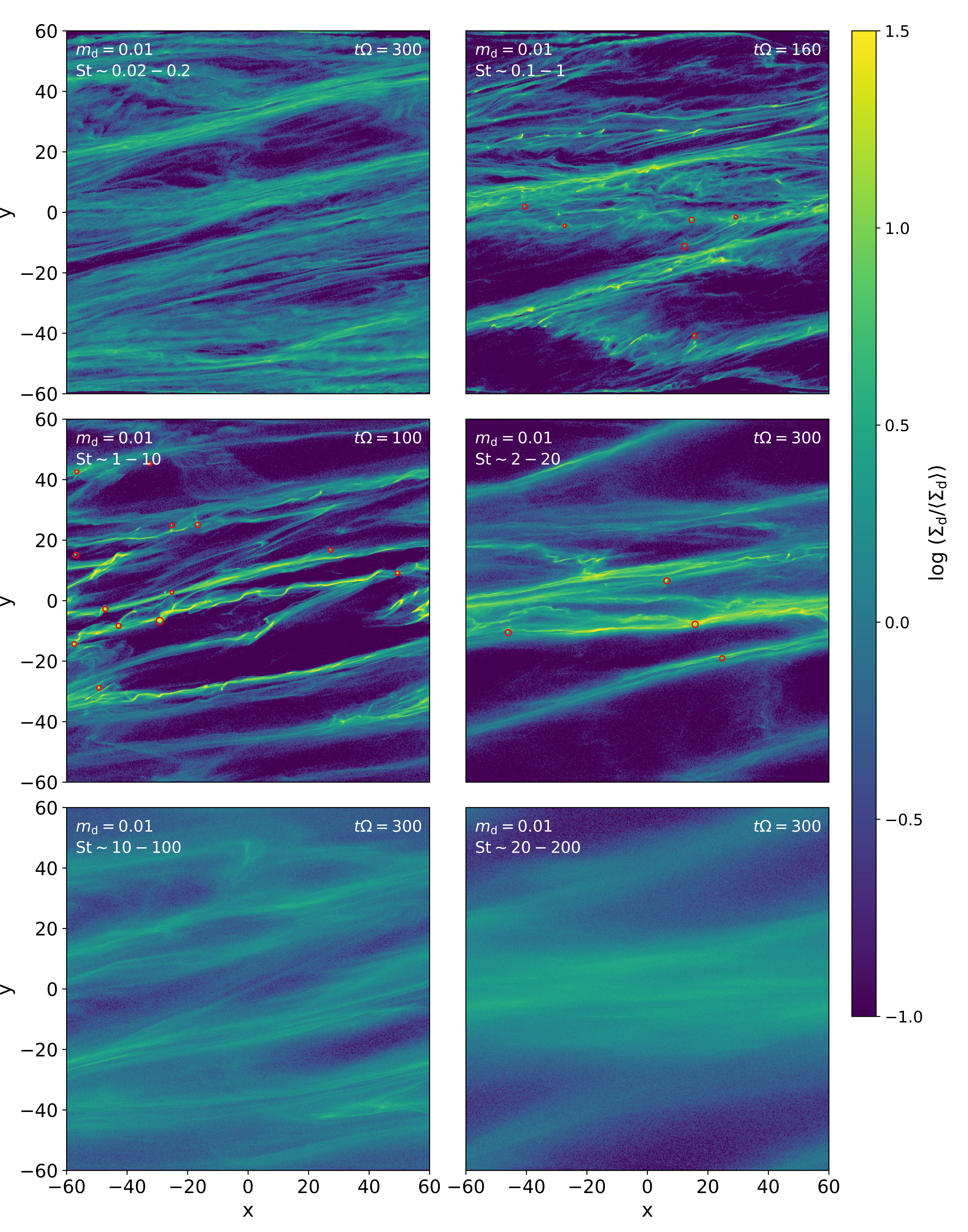}
    \caption{Suite of simulations with $m_{\rm d} = 0.01 m_{\rm g}$ and with $\mathrm{St} \sim 0.02 - 0.2$ (top left), $\mathrm{St} \sim 0.1 - 1$ (top right), $\mathrm{St} \sim 1 - 10$ (middle left), $\mathrm{St} \sim 2 - 20$ (middle right), $\mathrm{St} \sim 10 - 100$ (bottom left), and $\mathrm{St} \sim 20 - 200$ (bottom right).}
    \label{fig:dm1e-2_narr}
\end{figure*}

We also carried out a suite of additional simulations with $m_{\rm d} = 0.01 m_{\rm g}$ and with Stokes numbers in the range $\mathrm{St} \sim 0.02 - 0.2$, $\mathrm{St} \sim 0.1 - 1$, $\mathrm{St} \sim 1 - 10$, $\mathrm{St} \sim 2 - 20$, $\mathrm{St} \sim 10 - 100$, and $\mathrm{St} \sim 20 - 200$.  Figure \ref{fig:dm1e-2_narr} shows sample surface densities for these simulation.  These are typically at $t \Omega = 300$, except for the $\mathrm{St} \sim 0.1 - 1$ and $\mathrm{St} \sim 1 - 10$ simulations in which the high density clumps resulted in them being stopped at $t \Omega = 160$ and $t \Omega = 100$, respectively. 

There is no evidence of dense clumps forming in the $\mathrm{St} \sim 0.01 - 0.1$ (top left), $\mathrm{St} \sim 10 - 100$ (bottom left), or $\mathrm{St} \sim 20 - 200$ (bottom right) simulations.  However, dense clumps do form in the other 3 simulations.  This again indicates that when there is sufficient mass in particles with Stokes numbers close to unity, dense clumps are able to form in the particle disc.

It also seems that as long as there is sufficient mass in a small enough size range, dense clumps can form across a wide range of Stokes numbers, from $\mathrm{St} \sim 0.05$ up to $\mathrm{St} \sim 10$.  We should, however, be slightly cautious here since these are representative Stokes numbers based on the average midplane density and sound speed, and the actual Stokes numbers will depend on the local gas density and sound speed. 

\subsubsection{Criteria for collapse} \label{sec:collcrit}
As highlighted in \citet{baehr22}, a cloud of particles is unstable to collapse when its density is higher than the Hill density, given by
\begin{equation}
    \rho_{\rm Hill} = \frac{9}{4 \pi}\frac{M_*}{R^3},
\end{equation}
where $M_*$ is the mass of the central star and $R$ is the orbital radius of the particle cloud.  Since clumps form and live in the disc midplane, we analyze particle surface densities, rather than volume densities, and use:
\begin{equation}
    \Sigma_{\rm Hill} \sim 2 H_{\rm d} \rho_{\rm Hill} = \frac{9}{2 \pi} \frac{M_*}{R^3} H_{\rm d},  
    \label{eq:Sigma_Hill}
\end{equation}
where $H_{\rm d}$ is the scaleheight of the particle disc. If we use that $\Omega = \sqrt{G M_*/R^3} = 1$ and $G = 1$, we get that $\Sigma_{\rm Hill} \sim 9 H_{\rm d}/2 \pi$.  

As mentioned earlier, once the gas disc settles in a quasi-steady state, the scaleheight of the gas disc is $H_g \sim 5$.  Figures \ref{fig:dustscaleheight} and \ref{fig:refdust} both show that the scaleheight of the particle disc is substantially smaller than that of the gas disc, by about a factor of $4$ when considering the full size range ($\mathrm{St} \sim 0.02 - 200$). Vertical dust settling is even more noticeable when considering narrower size/Stokes number ranges, in particular for particles with Stokes numbers in the range $\mathrm{St} \sim 1$ to $\mathrm{St} \sim 10$.  

If we use $H_{\rm d} = 1$, then $\Sigma_{\rm Hill} \sim 9/2\pi \sim 1.4$.  Given an average gas surface density of $\Sigma_{\rm g} = 1$, this suggests that a local particle surface density comparable to the mean surface density of the gas disc would be sufficient for collapse.  

\begin{figure}
    \centering
    \includegraphics[width=0.49\textwidth]{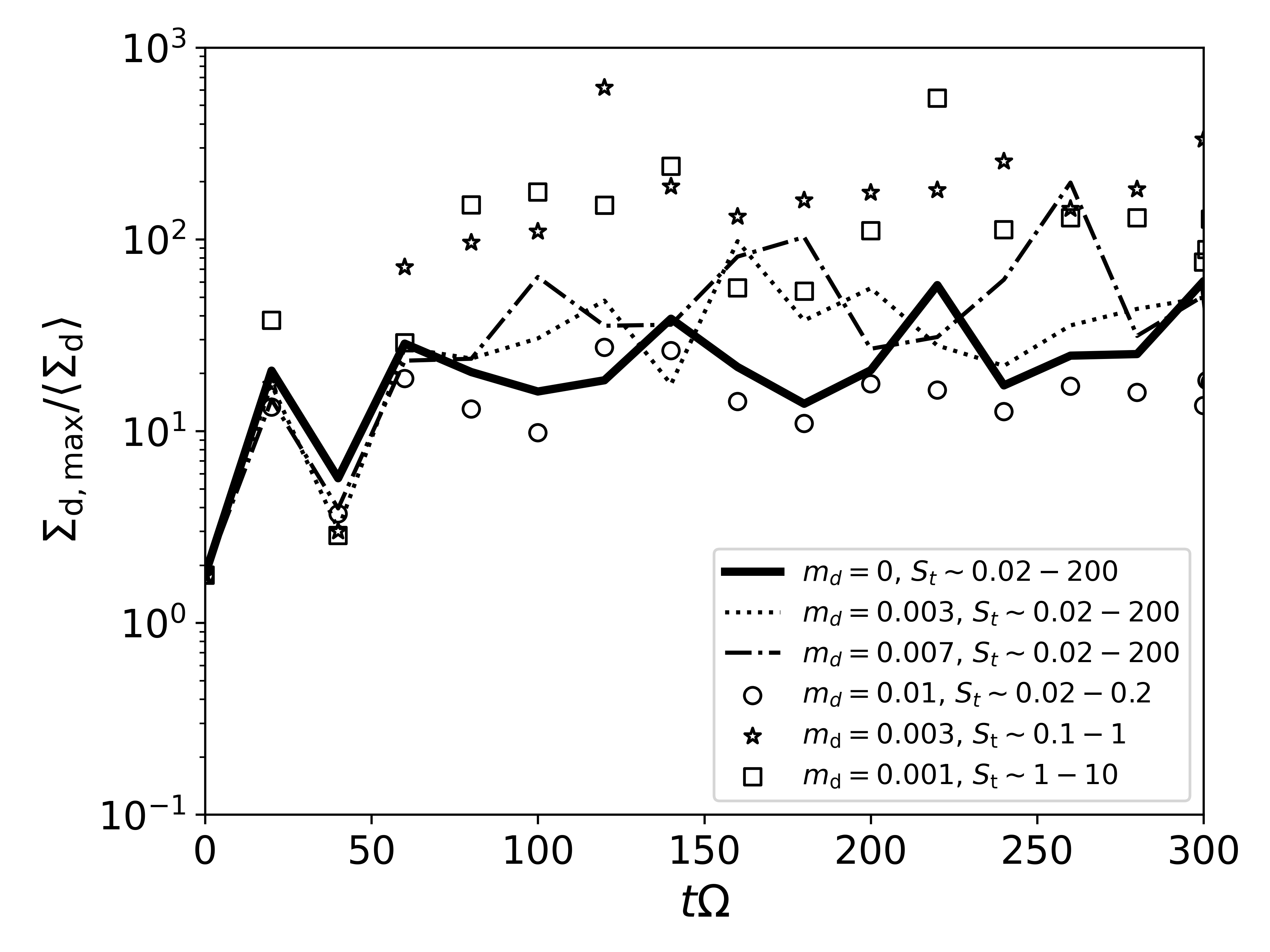}
    \includegraphics[width=0.49\textwidth]{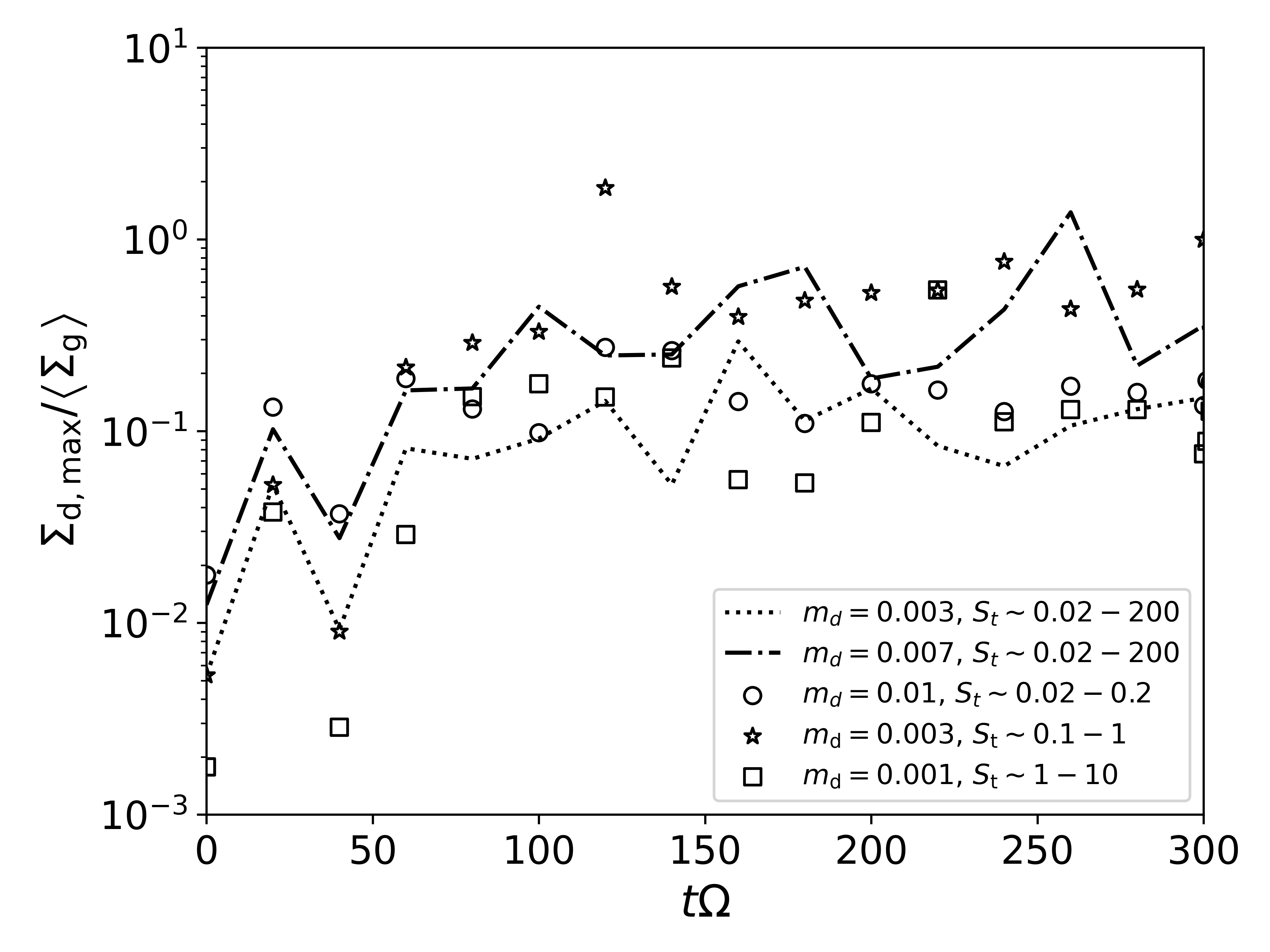}
    \caption{Dust surface density maxima, against time, in a sample of simulations in which dense particle clumps did not form. The lines are for simulations in which $\mathrm{St} \sim 0.02 - 200$, while the symbols are for three of the simulations with narrower Stokes number ranges. The top panel presents the particle surface densities relative to the mean particle surface density in that simulation, $\left< \Sigma_{\rm d} \right>$, and includes the test particle simulation (thick solid line). The bottom panel shows the simulations with massive particles with their surface densities plotted relative to the mean gas surface density, $\left< \Sigma_{\rm g} \right>$}
    \label{fig:sdens_max}
\end{figure}

Figure \ref{fig:sdens_max} shows the maximum particle surface density, against time, in a sample of the simulations that didn't produce dense clumps. The lines are simulations in which the Stokes numbers covered $\mathrm{St} \sim 0.02 - 200$, while the symbols are simulations in which the Stokes number range was narrower. The top panel shows the maximum surface density relative to the mean particle surface density in each simulation, and includes the test particle simulation (thick solid line). The bottom panel shows the simulations with massive particles, but with their surface densities normalised by the mean surface density of the gas disc.  

Figure \ref{fig:sdens_max} illustrates that when the Stokes number covers $\mathrm{St} \sim 0.02 - 200$ the maximum surface density will tend to be just over an order of magnitude greater than the mean, as also illustrated in Figure \ref{fig:sdensdistr}.  However, there is clearly also a lot of variability. In the $m_{\rm d} = 0.007 m_{\rm g}$, $\mathrm{St} \sim 0.02 - 200$ simulation (dash-dot line) there is even an epoch when the maximum surface density is about 200 times that of the mean.  This means that there will be regions where the particle surface density exceeds the mean gas surface (e.g., dash-dot line in the bottom panel of Figure \ref{fig:sdens_max}) and will be approaching values where collapse becomes possible.    

Figure \ref{fig:sdens_max} also shows that the maximum surface density will also depend on the range of Stokes numbers considered. When $\mathrm{St} \sim 0.1 - 1$ and $\mathrm{St} \sim 1 - 10$, the maximum particle surface density will typically be more than two orders of magnitude greater than the mean and can approach values of $\sim 1000$. Again, this suggests that if the total mass of particles with Stokes numbers around unity reaches about $10^{-3}$ that of the gas there can be regions of the disc where the local particle surface density becomes comparable to, or exceeds, the mean gas surface density (e.g., stars symbols in bottom panel of Figure \ref{fig:sdens_max}) and where it starts to become possible for these regions to gravitationally collapse to form dense clumps.

\subsubsection{Clump identification and properties} \label{sec:clumpid}

In order to quickly identify and analyse clumps that have formed in the simulations, we use a clump finding scheme based on the local density needed to remain bound, the Roche density. As shown by \citet{baehr22}, the Roche surface density, $\Sigma_R$, above which you might expect the particles to form bound clumps, can be approximated by:
\begin{equation}
    \Sigma_R = 8.8 \frac{\Omega^2 H_{\rm d}}{G},
    \label{eq:threshold}
\end{equation}
where $H_{\rm d}$ is again the scaleheight of the particle disc.  In these simulations, $\Omega = G = 1$ and, as mentioned in Section \ref{sec:collcrit}, the particle disc settles to a scaleheight of $H_{\rm d} \sim 1$, giving $\Sigma_R = 8.8$.  Since the average gas surface density is $\left< \Sigma_g \right> = 1$, if the particles have a total mass $1 \%$ that of the gas, then this would imply that bound clumps would only form in regions where the surface density has been enhanced by a factor of $8.8/0.01 = 880$.  

In the analysis presented here, we use $\Sigma_R = 10$, essentially identifying clumps as being locations where the local surface density of the solid particles is an order of magnitude greater than the average surface density of the gas. 

If we consider the $m_{\rm d} = 0.01 m_{\rm g}$ simulation with $\mathrm{St} \sim 0.02 - 200$ shown in the bottom-left panel in Figure \ref{fig:ptclsfullrange} there is one clump that satifies this criterion.  If we define the clump radius as being the Hill radius, then it contains $460014$ super-particles with a total mass of $6.5$ in code units. 

The actual mass of this clump will depend on the assumed scaling parameters.  If we use the same scaling as \citet{baehr22}, which is based on those in \citet{schafer17}, then the host star has a mass of $M_* = 1 M_\odot$ and the shearing box is assumed to be at $R = 50$~AU with a surface density of $\Sigma_{\rm o} = 53$~g~cm$^{-2}$, a temperature of $T = 11.25$~K, and a mean molecular weight of $\mu = 2.33$.  This gives an angular frequency of $\Omega = 5.63 \times 10^{-10}$~s$^{-1}$, a sound speed of $c_s = 199$~m~s$^{-1}$, and a scaleheight for the gas disc of $H_g' = c_s/\Omega = 3.53 \times 10^{11}$~m~$ = 2.36$~au. 

In code units the shearing box has $L_x = L_y = 120$ and $H_g = \pi$.  The total gas mass in the shearing box is then
\begin{equation}
M_{\rm tot,gas}' = \Sigma_{\rm o} \frac{L_x L_y}{\pi^2} H_g'^2= 9.36 \times 10^{31}~{\rm g} = 0.048~{\rm M}_\odot. 
\end{equation}
Given that the total gas mass in code units is $M_{\rm tot, gas} = L_x L_y$, the mass scale, $\hat{M}$, in the simulations is then 
\begin{equation}
\hat{M} = \frac{M_{\rm tot,gas}'}{M_{\rm tot,gas}} = \Sigma_{\rm o} \frac{H_g'^2}{\pi^2} = 6.69 \times 10^{27}~{\rm g} = 1.1~{\rm M_\oplus}.
\end{equation}
Hence, the clump in the $m_{\rm d} = 0.01 m_{\rm g}$ simulation mentioned above has a mass of $m_{\rm cl} = 6.5 \times 1.1~{\rm M_\oplus} = 7.15~{\rm M_\oplus}$. 

\begin{figure}
    \centering
    \includegraphics[width=0.49\textwidth]{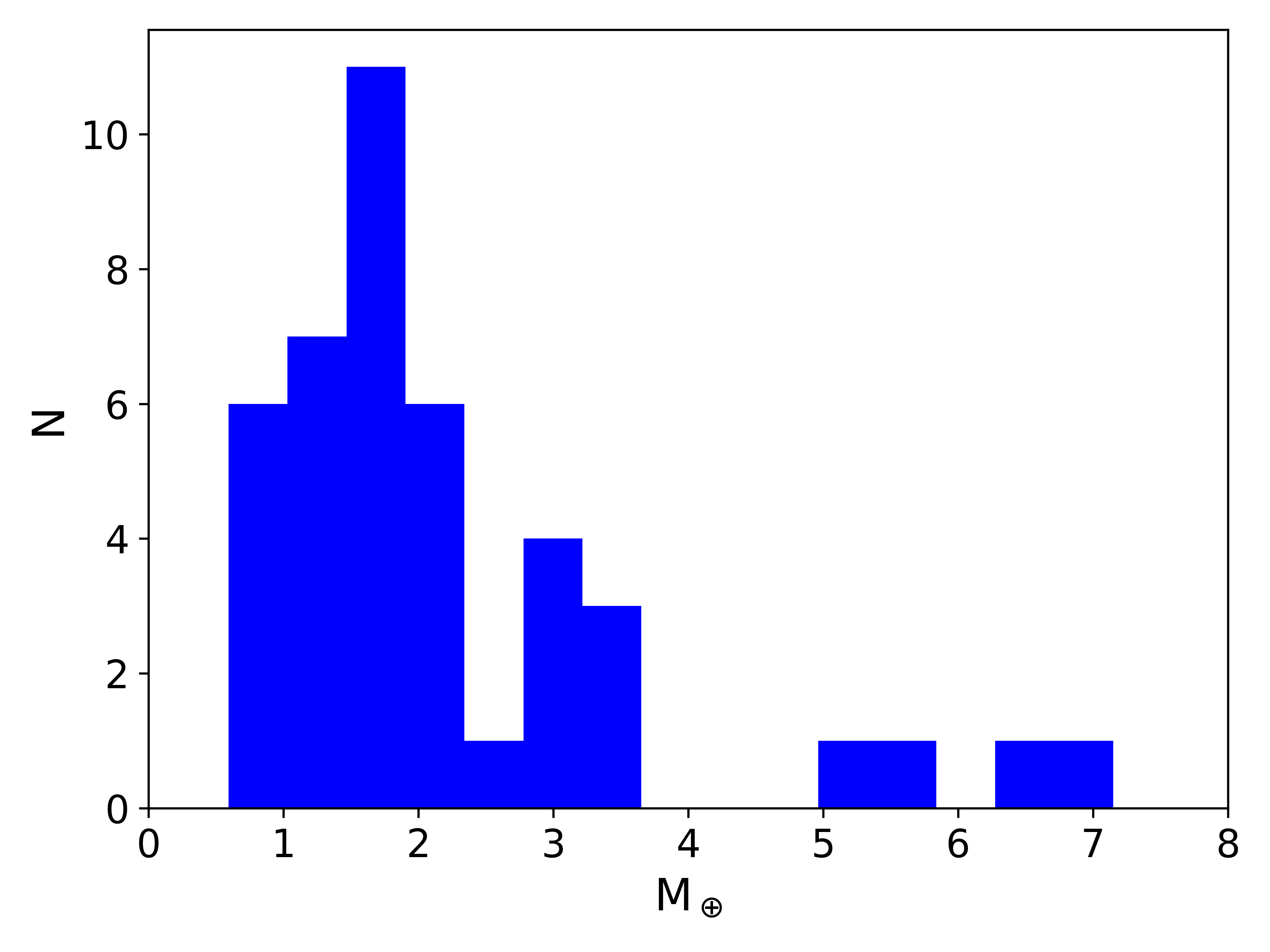}
    \caption{Histogram showing the distribution of clump masses that form in all that simulations in which dense clumps emerge (see Table \ref{tab:sims}). The masses are in $\mathrm{M}_\oplus$ based on the scaling discussed in Section \ref{sec:clumpid}.}
    \label{fig:clumpmasshist}
\end{figure}

If we consider all of the simulations that produced clumps (see Table \ref{tab:sims}), a total of 42 clumps formed that satisfied the criteria above.  The clump masses ranged from $m_{\rm cl} = 0.6~{\rm M_\oplus}$ to $m_{\rm cl} = 7.15~{\rm M_\oplus}$, with a mean of $m_{\rm cl} = 2.2~{\rm M_\oplus}$ and a standard deviation of $1.5~{\rm M_\oplus}$.  The distribution of clump masses is shown in Figure \ref{fig:clumpmasshist}.

Of course, this is not a homogeneous suite of simulations and the clump masses may depend on the dust-to-gas ratio and also on the range of Stokes numbers, which we aren't able to explore in more detail here.  However, the clump masses that emerge are consistent with the results presented in \citet{longarini23a,longarini23b} and \citet{rowther24a} which suggest that the gravitational collapse of the solid component of a two-fluid self-gravitating protostellar disc should produce planetary cores with masses between $1$ and $10$~M$_\oplus$. 

\begin{figure}
    \centering
    \includegraphics[width=0.95\linewidth]{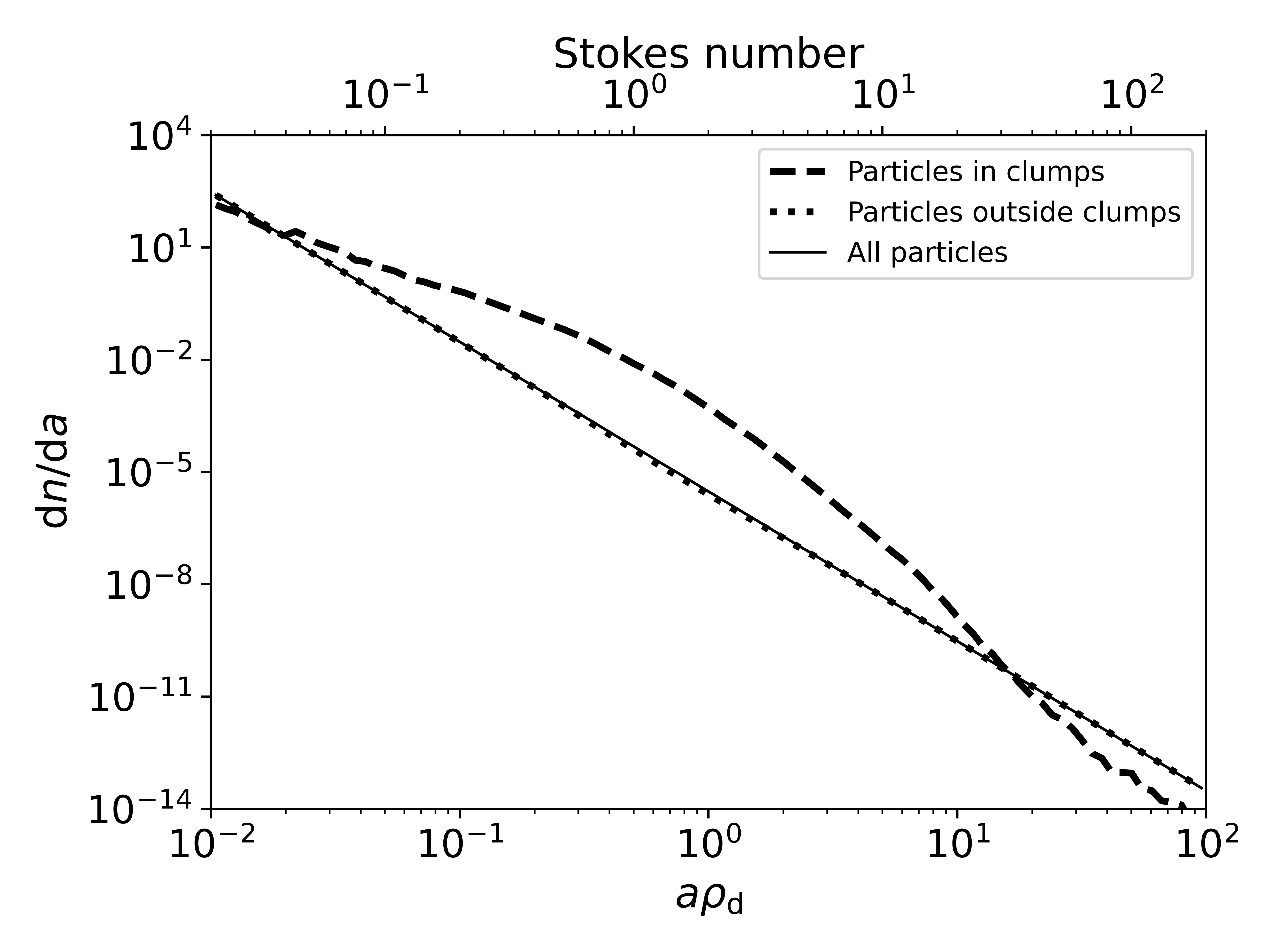}
    \caption{Size distribution for particles in clumps (dashed) line, not in clumps (dotted line) and all the particles (thin solid line) in the $m_{\rm d} = 0.025 m_{\rm g}$ simulation shown in the bottom right panel of Figure \ref{fig:ptclsfullrange}.}
    \label{fig:clumpsizedistr}
\end{figure}

Figure \ref{fig:clumpsizedistr} shows the size distribution of the particles in clumps (dashed line), the size distribution of particles not in clumps (dotted line), and the overall particle size distribution (thin solid line).  This is from the $m_{\rm d} = 0.025 m_{\rm g}$ simulation shown in the bottom panel of Figure \ref{fig:ptclsfullrange}, which formed 5 clumps by $t \Omega = 220$.  Each distribution in Figure \ref{fig:clumpsizedistr} has been normalised so that it integrates to 1 over the size range shown. The lower $x$-axis is $a \rho_d$, and corresponds to representative Stokes numbers ranging from $\mathrm{St} \sim 0.02$ to $\mathrm{St} \sim 200$ (see Equations \ref{eq:stoppingtime} and \ref{eq:Stokes}) which is shown in the upper x-axis.     

Figure \ref{fig:clumpsizedistr} shows that the clumps contain particles of all sizes, but that there is an overabundance of particles with Stokes numbers near unity.  This also slightly reduces the abundance of these particles in the rest of the disc (dotted line).  However, it is difficult to distinguish between the particle size distribution in the rest of the disc (dotted line) and the overall particle size distribution (thin solid line) because the total mass of particles in clumps in this simulation is only $4\%$ of the total particle mass.  

\subsubsection{Drag force assumption}
We can also use the scalings presented in Section \ref{sec:clumpid} to check our assumption that all the particles are in the Epstein drag regime. Our largest particles with Stokes numbers of $\mathrm{St} \sim 200$ will have sizes of $a = {200 c_s \rho_{\rm g}}/{\rho_{\rm d} \Omega} = 3400$~cm if $\rho_{\rm d} = 3$~g~cm$^{-3}$. The drag force is in the Epstein regime if $a < 9 \lambda / 4$, where $\lambda$ is the mean free path of the gas \citep[e.g.,][]{rice04}.  If we assume the gas is predominantly molecular hydrogen, then $\lambda = m_{\mathrm H_2}/A \rho_g$, where $A = 7 \times 10^{-16}$~cm$^2$ is the cross section of a hydrogen molecule.  Using the scalings presented in \ref{sec:clumpid} suggests that $\lambda \sim 15000$~cm and that the drag force would indeed be in the Epstein regime for all of the solid particles considered.

\section{Discussion} \label{sec:discussion}
It has long been known that in protostellar discs, density structures - such as self-gravitating spirals - can act to enhance the local density of solid particles, and that this is particularly effective for those particles that have Stokes numbers near unity \citep[e.g.,][]{rice04,gibbons12,booth2016,baehr21,longarini23a,rowther24a}. 

When dust does concentrate within the spiral structures of self-gravitating discs, it can potentially lead to the formation of gravitationally bound quantities of dust as large as the core of a planet, possibly suggesting another pathway for planet formation. Since previous studies have focused on single grain sizes with Stokes numbers near unity, they have assumed dust growth to be very efficient such that the entire mass budget is in these large grains.

Here we expand on this earlier work by considering particles with a range of sizes, rather than only considering a single particle size, or a single Stokes number.  The results indicate that even if the solid particles have a wide range of Stokes numbers, the local particle density can still be enhanced by more than an order of magnitude in the self-gravitating spirals, as long as the particle size distribution extends to $\mathrm{St} \sim 1$. 

In addition, if the total mass of particles is $\sim 1\%$ that of the gas and the particles have sizes that extend to $\mathrm{St} \sim 1$, the density enhancements can lead to the particle disc undergoing direct gravitational collapsing to form dense clumps.  When considering narrower size ranges, clumps can form as long as the particles span $\mathrm{St} \sim 0.5$ to $\mathrm{St} \sim 5$ and have a mass a few $\times 10^{-3}$ that of the gas.  There are also some indications that clumping may still occur even if the Stokes number range is slightly below unity (e.g., Figure \ref{fig:dm1e-2_narr}, where dense particle clumps emerged in the $\mathrm{St} \sim 0.1 - 1$ simulation).   

To provide some context, if we consider the scalings assumed in Section \ref{sec:clumpid}, then a $\mathrm{St} = 1$ particle at $R = 50$~au will have a size of $a = {c_s \rho_{\rm g}}/{\rho_{\rm d} \Omega} = 17$~cm if $\rho_{\rm d} = 3$~g~cm$^{-3}$. This suggests that at radii of 10s of au, where we expect this process might operate \citep{clarke09}, the presence of spiral density waves will significantly impact the concentration of solid particles if they have grown to $a \sim 1 - 10$~cm. 

If the particle size distribution extends from $a \sim 1$~$\mu$m to $a \sim 10$~cm, has  $n(a) \propto a^{-4}$, and if the solid particles have a total mass $1\%$ that of the gas, particles with sizes between $1$~cm and $10$~cm ($\mathrm{St} \sim 0.1$ to $\mathrm{St} \sim 1$) will have a mass $2 \times 10^{-3}$ that of the gas.  Our results would suggest that, in a self-gravitating disc with spiral density waves, a particle component with these properties would probably develop local particle density enhancements that could lead to the emergence of dense particle clumps.

Similarly, that the $n(a) \propto a^{-4}$ size distribution is slightly steeper than the canonical $n(a) \propto a^{-3.5}$ ISM grain size distribution might suggest that dense particle clumps are quite likely to form as long as the particles can grow to cm-sizes while the disc is still self-gravitating.

However, since a self-gravitating phase requires relatively high disc-to-star mass ratios \citep{kratter16} means that this phase will probably only persist for about $10^5$ years, or less \citep[e.g.,][]{hall19,rowther24b}. Hence, a key factor will be whether or not there is enough time for the solids to grow large enough while self-gravitating spirals are still present. Simple estimates of growth rates based upon collisions in a turbulent protoplanetary disc suggest growth can occur on time-scales of $1/(Z\Omega) \sim 10^{4}{\,\rm yr}$ at 50~au \citep{birnstiel}, which is independent of the strength of the turbulence due to the assumption that weaker turbulence leads to more settling and higher densities. 

However, in self-gravitating discs, both large-scale flows and the small scale-turbulence driven by secondary instabilities \citep{riols2017,booth2019} can counteract dust settling \citep[e.g.][]{riols2020} and the actual growth times may be longer. A more critical question is whether large dust grains can survive in the conditions present in self-gravitating discs since both the presence of shocks and strong relative motions driven by the turbulence may produce collision velocities that lead to fragmentation, rather than sticking. As in the simulations presented by \citep{booth2016}, we find that the smallest particles with $\mathrm{St} < 0.1$ can have velocity dispersions substantially smaller than the gas sound speed. Given our chosen scalings, this means that the collision velocity for these small particles may be of order $\sim 10\, {\rm m\,s^{-1}}$ which could allow icy solids to survive collisions \citep{booth2016}. 

There are, however, some indications that grain growth could actually start during the cloud collapse phase \citep[e.g.,][]{steinacker2015,bate22} and there is both observational \citep[e.g.,][]{galametz19,tychoniec20} and theoretical \citep[e.g.,][]{tu22,vorobyov24} evidence for grain growth to $\sim$mm-sizes in very young protostellar discs, suggesting that planet formation may start during the earliest stages of star formation \citep{harsono18}.  If the enhancement of the solids can then lead to the direct formation of planetesimals, or even planetary-mass cores, that would overcome many of the barriers to planet formation via core accretion \citep{nixon18}.

\section{Conclusion} \label{sec:conclusion}
The results presented here indicate that the presence of self-gravitating spirals can lead to substantial enhancements in solid grain concentrations even if the particles have a wide range of sizes.  It does, however, still require that the particle size distribution extends to particles that have Stokes numbers close to, or even above, unity.  

The results also indicate that if there is enough mass in the solid particles, the density enhancements can be sufficient for the particle disc to gravitationally collapse to directly form dense clumps with up to a few M$_{\oplus}$ of solid material. Therefore, these could form super-Earths, or the cores of future giant planets that then grow through accretion of gas from the disc. The self-gravitating phase is expected to occur in the first $\sim10^5$~years of the disc lifetime so, assuming clumps form in this time, we can generate planetary cores in the Class 0 phase as required by observations of accreting protoplanets in Class I discs. Additionally, even when bound clumps of solids do not form, the enhancements of dust density in the spiral waves could assist grain growth, as long as the collision velocities of the smallest particles is low enough to allow collisions to lead to growth, rather than fragmentation.

What's particularly interesting is that these results suggest that clumping requires a dust-to-gas ratio of $\sim 0.01$, close to what is expected in very young protostellar systems.  Furthermore, if we consider the particles with Stokes numbers between $\mathrm{St} \sim 1$ and $\mathrm{St} \sim 10$, the required dust-to-gas ratio is substantially smaller than 0.01, with dense clumps emerging for dust-to-gas ratios as low as $m_{\rm d}/m_{\rm g} = 0.003$.  Hence, the conditions required for clumping are at least consistent with what might be expected in these very young systems.

There are, of course, caveats.  A key issue is that the solid particles need to grow to sizes such that the Stokes number is $\mathrm{St} \sim 1$.  At orbital radii where we might expect self-gravity to manifest as spiral density waves (a few 10s of au) this would require the particles to grow to $a \sim 10$~cm and would need to do so while the system is still very young. 

Therefore, it's not clear that grain growth can be rapid enough for particles to grow to these sizes during the epoch when the system is expected to be self-gravitating (within the first ~$\sim 10^5$~years). However, since the spirals will start to concentrate the grains at Stokes numbers well below unity, this could itself contribute to accelerated grain growth at these early times. 

Observations of very young protostellar stystems at sub-mm or radio wavelengths that either indicate the presence of mm- to cm-sized grains, or rule them out, would be a powerful probe of whether or not the mechanism proposed here could actually operate. What is intriguing, though, is that there is increasing evidence that planet formation needs to start during the earliest phases of protostellar evolution, the same epoch as the process described here is expected to operate.  This would therefore seem to be a mechanism that is worth exploring in more detail.

\section*{Acknowledgements}
KR and AKY acknowledge support from the UK Science and
Technologies Facilities Council (STFC) via grant ST/V000594/1. RAB thanks the Royal Society for their support in the form of a University Research Fellowship. SR acknowledges support funding from the Science \& Technology Facilities Council (STFC) through Consolidated Grant ST/W000857/1. FM acknowledges support from the Royal Society Dorothy Hodgkin Fellowship. This work used the DiRAC Data Intensive service (DIaL2) at the University of Leicester, managed by the University of Leicester Research Computing Service on behalf of the STFC DiRAC HPC Facility (www.dirac.ac.uk). The DiRAC service at Leicester was funded by BEIS, UKRI and STFC capital funding and STFC operations grants. DiRAC is part of the UKRI Digital Research Infrastructure. For the purpose of open access, the author has applied a Creative Commons Attribution (CC BY) licence to any Author Accepted Manuscript version arising from this submission.

\section*{Data Availability}
All of the raw data files, processed data files and python scripts used to generate the figures presented in the paper will be archived will be publicly available in the Research Data Management Zenodo repository at \url{https://doi.org/10.5281/zenodo.15280731}.  Additional data files such as initial data files and data files for intermediate steps in the simulations will be stored locally and made available on request. The {\sc pencil} code is available at \url{https://github.com/pencil-code}.



\bibliographystyle{mnras}
\bibliography{shearing_particles} 

\begin{thebibliography}{}
\makeatletter
\relax
\def\mn@urlcharsother{\let\do\@makeother \do\$\do\&\do\#\do\^\do\_\do\%\do\~}
\def\mn@doi{\begingroup\mn@urlcharsother \@ifnextchar [ {\mn@doi@}
  {\mn@doi@[]}}
\def\mn@doi@[#1]#2{\def\@tempa{#1}\ifx\@tempa\@empty \href
  {http://dx.doi.org/#2} {doi:#2}\else \href {http://dx.doi.org/#2} {#1}\fi
  \endgroup}
\def\mn@eprint#1#2{\mn@eprint@#1:#2::\@nil}
\def\mn@eprint@arXiv#1{\href {http://arxiv.org/abs/#1} {{\tt arXiv:#1}}}
\def\mn@eprint@dblp#1{\href {http://dblp.uni-trier.de/rec/bibtex/#1.xml}
  {dblp:#1}}
\def\mn@eprint@#1:#2:#3:#4\@nil{\def\@tempa {#1}\def\@tempb {#2}\def\@tempc
  {#3}\ifx \@tempc \@empty \let \@tempc \@tempb \let \@tempb \@tempa \fi \ifx
  \@tempb \@empty \def\@tempb {arXiv}\fi \@ifundefined
  {mn@eprint@\@tempb}{\@tempb:\@tempc}{\expandafter \expandafter \csname
  mn@eprint@\@tempb\endcsname \expandafter{\@tempc}}}

\bibitem[\protect\citeauthoryear{{Baehr} \& {Zhu}}{{Baehr} \&
  {Zhu}}{2021}]{baehr21}
{Baehr} H.,  {Zhu} Z.,  2021, \mn@doi [\apj] {10.3847/1538-4357/abddb3}, \href
  {https://ui.adsabs.harvard.edu/abs/2021ApJ...909..135B} {909, 135}

\bibitem[\protect\citeauthoryear{{Baehr}, {Klahr}  \& {Kratter}}{{Baehr}
  et~al.}{2017}]{baehr17}
{Baehr} H.,  {Klahr} H.,   {Kratter} K.~M.,  2017, \mn@doi [\apj]
  {10.3847/1538-4357/aa8a66}, \href
  {https://ui.adsabs.harvard.edu/abs/2017ApJ...848...40B} {848, 40}

\bibitem[\protect\citeauthoryear{{Baehr}, {Zhu}  \& {Yang}}{{Baehr}
  et~al.}{2022}]{baehr22}
{Baehr} H.,  {Zhu} Z.,   {Yang} C.-C.,  2022, \mn@doi [\apj]
  {10.3847/1538-4357/ac7228}, \href
  {https://ui.adsabs.harvard.edu/abs/2022ApJ...933..100B} {933, 100}

\bibitem[\protect\citeauthoryear{{Bai} \& {Stone}}{{Bai} \&
  {Stone}}{2010}]{bai10}
{Bai} X.-N.,  {Stone} J.~M.,  2010, \mn@doi [\apj]
  {10.1088/0004-637X/722/2/1437}, \href
  {https://ui.adsabs.harvard.edu/abs/2010ApJ...722.1437B} {722, 1437}

\bibitem[\protect\citeauthoryear{{Bate}}{{Bate}}{2022}]{bate22}
{Bate} M.~R.,  2022, \mn@doi [\mnras] {10.1093/mnras/stac1391}, \href
  {https://ui.adsabs.harvard.edu/abs/2022MNRAS.514.2145B} {514, 2145}

\bibitem[\protect\citeauthoryear{{Birnstiel}, {Klahr}  \&
  {Ercolano}}{{Birnstiel} et~al.}{2012}]{birnstiel}
{Birnstiel} T.,  {Klahr} H.,   {Ercolano} B.,  2012, \mn@doi [\aap]
  {10.1051/0004-6361/201118136}, \href
  {https://ui.adsabs.harvard.edu/abs/2012A&A...539A.148B} {539, A148}

\bibitem[\protect\citeauthoryear{{Blum} \& {Wurm}}{{Blum} \&
  {Wurm}}{2008}]{blum08}
{Blum} J.,  {Wurm} G.,  2008, \mn@doi [\araa]
  {10.1146/annurev.astro.46.060407.145152}, \href
  {https://ui.adsabs.harvard.edu/abs/2008ARA&A..46...21B} {46, 21}

\bibitem[\protect\citeauthoryear{{Booth} \& {Clarke}}{{Booth} \&
  {Clarke}}{2016}]{booth2016}
{Booth} R.~A.,  {Clarke} C.~J.,  2016, \mn@doi [\mnras] {10.1093/mnras/stw488},
  \href {https://ui.adsabs.harvard.edu/abs/2016MNRAS.458.2676B} {458, 2676}

\bibitem[\protect\citeauthoryear{{Booth} \& {Clarke}}{{Booth} \&
  {Clarke}}{2019}]{booth2019}
{Booth} R.~A.,  {Clarke} C.~J.,  2019, \mn@doi [\mnras]
  {10.1093/mnras/sty3340}, \href
  {https://ui.adsabs.harvard.edu/abs/2019MNRAS.483.3718B} {483, 3718}

\bibitem[\protect\citeauthoryear{{Boss}}{{Boss}}{1997}]{boss97}
{Boss} A.~P.,  1997, \mn@doi [Science] {10.1126/science.276.5320.1836}, \href
  {https://ui.adsabs.harvard.edu/abs/1997Sci...276.1836B} {276, 1836}

\bibitem[\protect\citeauthoryear{{Brandenburg}}{{Brandenburg}}{2003}]{brandenburg03}
{Brandenburg} A.,  2003, in {Ferriz-Mas} A.,  {N{\'u}{\~n}ez} M.,  eds, ,
  Advances in Nonlinear Dynamics.
p.~269, \mn@doi{10.1201/9780203493137.ch9}

\bibitem[\protect\citeauthoryear{{Clarke} \& {Lodato}}{{Clarke} \&
  {Lodato}}{2009}]{clarke09}
{Clarke} C.~J.,  {Lodato} G.,  2009, \mn@doi [\mnras]
  {10.1111/j.1745-3933.2009.00695.x}, \href
  {https://ui.adsabs.harvard.edu/abs/2009MNRAS.398L...6C} {398, L6}

\bibitem[\protect\citeauthoryear{{Durisen}, {Boss}, {Mayer}, {Nelson}, {Quinn}
  \& {Rice}}{{Durisen} et~al.}{2007}]{durisen07}
{Durisen} R.~H.,  {Boss} A.~P.,  {Mayer} L.,  {Nelson} A.~F.,  {Quinn} T.,
  {Rice} W.~K.~M.,  2007, in {Reipurth} B.,  {Jewitt} D.,   {Keil} K.,  eds,
  Protostars and Planets V. p.~607 (\mn@eprint {arXiv} {astro-ph/0603179}),
  \mn@doi{10.48550/arXiv.astro-ph/0603179}

\bibitem[\protect\citeauthoryear{{Forgan}, {Hall}, {Meru}  \& {Rice}}{{Forgan}
  et~al.}{2018}]{forgan18}
{Forgan} D.~H.,  {Hall} C.,  {Meru} F.,   {Rice} W.~K.~M.,  2018, \mn@doi
  [\mnras] {10.1093/mnras/stx2870}, \href
  {https://ui.adsabs.harvard.edu/abs/2018MNRAS.474.5036F} {474, 5036}

\bibitem[\protect\citeauthoryear{{Galametz}, {Maury}, {Valdivia}, {Testi},
  {Belloche}  \& {Andr{\'e}}}{{Galametz} et~al.}{2019}]{galametz19}
{Galametz} M.,  {Maury} A.~J.,  {Valdivia} V.,  {Testi} L.,  {Belloche} A.,
  {Andr{\'e}} P.,  2019, \mn@doi [\aap] {10.1051/0004-6361/201936342}, \href
  {https://ui.adsabs.harvard.edu/abs/2019A&A...632A...5G} {632, A5}

\bibitem[\protect\citeauthoryear{{Gammie}}{{Gammie}}{2001}]{gammie01}
{Gammie} C.~F.,  2001, \mn@doi [\apj] {10.1086/320631}, \href
  {https://ui.adsabs.harvard.edu/abs/2001ApJ...553..174G} {553, 174}

\bibitem[\protect\citeauthoryear{{Gibbons}, {Rice}  \&
  {Mamatsashvili}}{{Gibbons} et~al.}{2012}]{gibbons12}
{Gibbons} P.~G.,  {Rice} W.~K.~M.,   {Mamatsashvili} G.~R.,  2012, \mn@doi
  [\mnras] {10.1111/j.1365-2966.2012.21731.x}, \href
  {https://ui.adsabs.harvard.edu/abs/2012MNRAS.426.1444G} {426, 1444}

\bibitem[\protect\citeauthoryear{{Gibbons}, {Mamatsashvili}  \&
  {Rice}}{{Gibbons} et~al.}{2014}]{gibbons14}
{Gibbons} P.~G.,  {Mamatsashvili} G.~R.,   {Rice} W.~K.~M.,  2014, \mn@doi
  [\mnras] {10.1093/mnras/stu809}, \href
  {https://ui.adsabs.harvard.edu/abs/2014MNRAS.442..361G} {442, 361}

\bibitem[\protect\citeauthoryear{{Gibbons}, {Mamatsashvili}  \&
  {Rice}}{{Gibbons} et~al.}{2015}]{gibbons15}
{Gibbons} P.~G.,  {Mamatsashvili} G.~R.,   {Rice} W.~K.~M.,  2015, \mn@doi
  [\mnras] {10.1093/mnras/stv1766}, \href
  {https://ui.adsabs.harvard.edu/abs/2015MNRAS.453.4232G} {453, 4232}

\bibitem[\protect\citeauthoryear{{Godon} \& {Livio}}{{Godon} \&
  {Livio}}{1999}]{godon99}
{Godon} P.,  {Livio} M.,  1999, \mn@doi [\apj] {10.1086/307720}, \href
  {https://ui.adsabs.harvard.edu/abs/1999ApJ...523..350G} {523, 350}

\bibitem[\protect\citeauthoryear{{Haghighipour} \& {Boss}}{{Haghighipour} \&
  {Boss}}{2003}]{haghighipour03}
{Haghighipour} N.,  {Boss} A.~P.,  2003, \mn@doi [\apj] {10.1086/378950}, \href
  {https://ui.adsabs.harvard.edu/abs/2003ApJ...598.1301H} {598, 1301}

\bibitem[\protect\citeauthoryear{{Hall}, {Dong}, {Rice}, {Harries}, {Najita},
  {Alexander}  \& {Brittain}}{{Hall} et~al.}{2019}]{hall19}
{Hall} C.,  {Dong} R.,  {Rice} K.,  {Harries} T.~J.,  {Najita} J.,  {Alexander}
  R.,   {Brittain} S.,  2019, \mn@doi [\apj] {10.3847/1538-4357/aafac2}, \href
  {https://ui.adsabs.harvard.edu/abs/2019ApJ...871..228H} {871, 228}

\bibitem[\protect\citeauthoryear{{Harsono}, {Bjerkeli}, {van der Wiel},
  {Ramsey}, {Maud}, {Kristensen}  \& {J{\o}rgensen}}{{Harsono}
  et~al.}{2018}]{harsono18}
{Harsono} D.,  {Bjerkeli} P.,  {van der Wiel} M. H.~D.,  {Ramsey} J.~P.,
  {Maud} L.~T.,  {Kristensen} L.~E.,   {J{\o}rgensen} J.~K.,  2018, \mn@doi
  [Nature Astronomy] {10.1038/s41550-018-0497-x}, \href
  {https://ui.adsabs.harvard.edu/abs/2018NatAs...2..646H} {2, 646}

\bibitem[\protect\citeauthoryear{{Johansen} \& {Youdin}}{{Johansen} \&
  {Youdin}}{2007}]{johansen07}
{Johansen} A.,  {Youdin} A.,  2007, \mn@doi [\apj] {10.1086/516730}, \href
  {https://ui.adsabs.harvard.edu/abs/2007ApJ...662..627J} {662, 627}

\bibitem[\protect\citeauthoryear{{Johansen}, {Andersen}  \&
  {Brandenburg}}{{Johansen} et~al.}{2004}]{johansen04}
{Johansen} A.,  {Andersen} A.~C.,   {Brandenburg} A.,  2004, \mn@doi [\aap]
  {10.1051/0004-6361:20034417}, \href
  {https://ui.adsabs.harvard.edu/abs/2004A&A...417..361J} {417, 361}

\bibitem[\protect\citeauthoryear{{Johansen}, {Klahr}  \& {Henning}}{{Johansen}
  et~al.}{2006}]{johansen06}
{Johansen} A.,  {Klahr} H.,   {Henning} T.,  2006, \mn@doi [\apj]
  {10.1086/498078}, \href
  {https://ui.adsabs.harvard.edu/abs/2006ApJ...636.1121J} {636, 1121}

\bibitem[\protect\citeauthoryear{{Johansen}, {Youdin}  \& {Mac Low}}{{Johansen}
  et~al.}{2009}]{johansen09}
{Johansen} A.,  {Youdin} A.,   {Mac Low} M.-M.,  2009, \mn@doi [\apjl]
  {10.1088/0004-637X/704/2/L75}, \href
  {https://ui.adsabs.harvard.edu/abs/2009ApJ...704L..75J} {704, L75}

\bibitem[\protect\citeauthoryear{{Klahr} \& {Schreiber}}{{Klahr} \&
  {Schreiber}}{2020}]{klahr20}
{Klahr} H.,  {Schreiber} A.,  2020, \mn@doi [\apj] {10.3847/1538-4357/abac58},
  \href {https://ui.adsabs.harvard.edu/abs/2020ApJ...901...54K} {901, 54}

\bibitem[\protect\citeauthoryear{{Kratter} \& {Lodato}}{{Kratter} \&
  {Lodato}}{2016}]{kratter16}
{Kratter} K.,  {Lodato} G.,  2016, \mn@doi [\araa]
  {10.1146/annurev-astro-081915-023307}, \href
  {https://ui.adsabs.harvard.edu/abs/2016ARA&A..54..271K} {54, 271}

\bibitem[\protect\citeauthoryear{{Kratter}, {Murray-Clay}  \&
  {Youdin}}{{Kratter} et~al.}{2010}]{kratter10}
{Kratter} K.~M.,  {Murray-Clay} R.~A.,   {Youdin} A.~N.,  2010, \mn@doi [\apj]
  {10.1088/0004-637X/710/2/1375}, \href
  {https://ui.adsabs.harvard.edu/abs/2010ApJ...710.1375K} {710, 1375}

\bibitem[\protect\citeauthoryear{{Kuiper}}{{Kuiper}}{1951}]{kuiper51}
{Kuiper} G.~P.,  1951, \mn@doi [Proceedings of the National Academy of Science]
  {10.1073/pnas.37.1.1}, \href
  {https://ui.adsabs.harvard.edu/abs/1951PNAS...37....1K} {37, 1}

\bibitem[\protect\citeauthoryear{{Laibe}}{{Laibe}}{2014}]{laibe14}
{Laibe} G.,  2014, \mn@doi [\mnras] {10.1093/mnras/stt1928}, \href
  {https://ui.adsabs.harvard.edu/abs/2014MNRAS.437.3037L} {437, 3037}

\bibitem[\protect\citeauthoryear{{Lin} \& {Pringle}}{{Lin} \&
  {Pringle}}{1990}]{lin90}
{Lin} D. N.~C.,  {Pringle} J.~E.,  1990, \mn@doi [\apj] {10.1086/169004}, \href
  {https://ui.adsabs.harvard.edu/abs/1990ApJ...358..515L} {358, 515}

\bibitem[\protect\citeauthoryear{{Lodato} \& {Rice}}{{Lodato} \&
  {Rice}}{2004}]{lodato04}
{Lodato} G.,  {Rice} W.~K.~M.,  2004, \mn@doi [\mnras]
  {10.1111/j.1365-2966.2004.07811.x}, \href
  {https://ui.adsabs.harvard.edu/abs/2004MNRAS.351..630L} {351, 630}

\bibitem[\protect\citeauthoryear{{Longarini}, {Lodato}, {Bertin}  \&
  {Armitage}}{{Longarini} et~al.}{2023a}]{longarini23a}
{Longarini} C.,  {Lodato} G.,  {Bertin} G.,   {Armitage} P.~J.,  2023a, \mn@doi
  [\mnras] {10.1093/mnras/stac3653}, \href
  {https://ui.adsabs.harvard.edu/abs/2023MNRAS.519.2017L} {519, 2017}

\bibitem[\protect\citeauthoryear{{Longarini}, {Armitage}, {Lodato}, {Price}  \&
  {Ceppi}}{{Longarini} et~al.}{2023b}]{longarini23b}
{Longarini} C.,  {Armitage} P.~J.,  {Lodato} G.,  {Price} D.~J.,   {Ceppi} S.,
  2023b, \mn@doi [\mnras] {10.1093/mnras/stad1400}, \href
  {https://ui.adsabs.harvard.edu/abs/2023MNRAS.522.6217L} {522, 6217}

\bibitem[\protect\citeauthoryear{{Mathis}, {Rumpl}  \& {Nordsieck}}{{Mathis}
  et~al.}{1977}]{mathis77}
{Mathis} J.~S.,  {Rumpl} W.,   {Nordsieck} K.~H.,  1977, \mn@doi [\apj]
  {10.1086/155591}, \href
  {https://ui.adsabs.harvard.edu/abs/1977ApJ...217..425M} {217, 425}

\bibitem[\protect\citeauthoryear{{Mizuno}}{{Mizuno}}{1980}]{mizuno80}
{Mizuno} H.,  1980, \mn@doi [Progress of Theoretical Physics]
  {10.1143/PTP.64.544}, \href
  {https://ui.adsabs.harvard.edu/abs/1980PThPh..64..544M} {64, 544}

\bibitem[\protect\citeauthoryear{{M{\"u}ller}, {Helled}  \&
  {Mayer}}{{M{\"u}ller} et~al.}{2018}]{muller18}
{M{\"u}ller} S.,  {Helled} R.,   {Mayer} L.,  2018, \mn@doi [\apj]
  {10.3847/1538-4357/aaa840}, \href
  {https://ui.adsabs.harvard.edu/abs/2018ApJ...854..112M} {854, 112}

\bibitem[\protect\citeauthoryear{{Nero} \& {Bjorkman}}{{Nero} \&
  {Bjorkman}}{2009}]{nero09}
{Nero} D.,  {Bjorkman} J.~E.,  2009, \mn@doi [\apjl]
  {10.1088/0004-637X/702/2/L163}, \href
  {https://ui.adsabs.harvard.edu/abs/2009ApJ...702L.163N} {702, L163}

\bibitem[\protect\citeauthoryear{{Nixon}, {King}  \& {Pringle}}{{Nixon}
  et~al.}{2018}]{nixon18}
{Nixon} C.~J.,  {King} A.~R.,   {Pringle} J.~E.,  2018, \mn@doi [\mnras]
  {10.1093/mnras/sty593}, \href
  {https://ui.adsabs.harvard.edu/abs/2018MNRAS.477.3273N} {477, 3273}

\bibitem[\protect\citeauthoryear{{Pencil Code Collaboration} et~al.,}{{Pencil
  Code Collaboration} et~al.}{2021}]{pencilcodecoll}
{Pencil Code Collaboration} et~al., 2021, \mn@doi [The Journal of Open Source
  Software] {10.21105/joss.02807}, \href
  {https://ui.adsabs.harvard.edu/abs/2021JOSS....6.2807P} {6, 2807}

\bibitem[\protect\citeauthoryear{{Pollack}, {Hubickyj}, {Bodenheimer},
  {Lissauer}, {Podolak}  \& {Greenzweig}}{{Pollack} et~al.}{1996}]{pollack96}
{Pollack} J.~B.,  {Hubickyj} O.,  {Bodenheimer} P.,  {Lissauer} J.~J.,
  {Podolak} M.,   {Greenzweig} Y.,  1996, \mn@doi [\icarus]
  {10.1006/icar.1996.0190}, \href
  {https://ui.adsabs.harvard.edu/abs/1996Icar..124...62P} {124, 62}

\bibitem[\protect\citeauthoryear{{Rice}, {Lodato}, {Pringle}, {Armitage}  \&
  {Bonnell}}{{Rice} et~al.}{2004}]{rice04}
{Rice} W.~K.~M.,  {Lodato} G.,  {Pringle} J.~E.,  {Armitage} P.~J.,   {Bonnell}
  I.~A.,  2004, \mn@doi [\mnras] {10.1111/j.1365-2966.2004.08339.x}, \href
  {https://ui.adsabs.harvard.edu/abs/2004MNRAS.355..543R} {355, 543}

\bibitem[\protect\citeauthoryear{{Rice}, {Lodato}  \& {Armitage}}{{Rice}
  et~al.}{2005}]{rice05}
{Rice} W.~K.~M.,  {Lodato} G.,   {Armitage} P.~J.,  2005, \mn@doi [\mnras]
  {10.1111/j.1745-3933.2005.00105.x}, \href
  {https://ui.adsabs.harvard.edu/abs/2005MNRAS.364L..56R} {364, L56}

\bibitem[\protect\citeauthoryear{{Rice}, {Lodato}, {Pringle}, {Armitage}  \&
  {Bonnell}}{{Rice} et~al.}{2006}]{rice06}
{Rice} W.~K.~M.,  {Lodato} G.,  {Pringle} J.~E.,  {Armitage} P.~J.,   {Bonnell}
  I.~A.,  2006, \mn@doi [\mnras] {10.1111/j.1745-3933.2006.00215.x}, \href
  {https://ui.adsabs.harvard.edu/abs/2006MNRAS.372L...9R} {372, L9}

\bibitem[\protect\citeauthoryear{{Rice}, {Mayo}  \& {Armitage}}{{Rice}
  et~al.}{2010}]{rice10}
{Rice} W.~K.~M.,  {Mayo} J.~H.,   {Armitage} P.~J.,  2010, \mn@doi [\mnras]
  {10.1111/j.1365-2966.2009.15992.x}, \href
  {https://ui.adsabs.harvard.edu/abs/2010MNRAS.402.1740R} {402, 1740}

\bibitem[\protect\citeauthoryear{{Rice}, {Lopez}, {Forgan}  \& {Biller}}{{Rice}
  et~al.}{2015}]{rice15}
{Rice} K.,  {Lopez} E.,  {Forgan} D.,   {Biller} B.,  2015, \mn@doi [\mnras]
  {10.1093/mnras/stv1997}, \href
  {https://ui.adsabs.harvard.edu/abs/2015MNRAS.454.1940R} {454, 1940}

\bibitem[\protect\citeauthoryear{{Riols}, {Latter}  \& {Paardekooper}}{{Riols}
  et~al.}{2017}]{riols2017}
{Riols} A.,  {Latter} H.,   {Paardekooper} S.~J.,  2017, \mn@doi [\mnras]
  {10.1093/mnras/stx1548}, \href
  {https://ui.adsabs.harvard.edu/abs/2017MNRAS.471..317R} {471, 317}

\bibitem[\protect\citeauthoryear{{Riols}, {Roux}, {Latter}  \& {Lesur}}{{Riols}
  et~al.}{2020}]{riols2020}
{Riols} A.,  {Roux} B.,  {Latter} H.,   {Lesur} G.,  2020, \mn@doi [\mnras]
  {10.1093/mnras/staa567}, \href
  {https://ui.adsabs.harvard.edu/abs/2020MNRAS.493.4631R} {493, 4631}

\bibitem[\protect\citeauthoryear{{Rowther}, {Nealon}, {Meru}, {Wurster}, {Aly},
  {Alexander}, {Rice}  \& {Booth}}{{Rowther} et~al.}{2024a}]{rowther24a}
{Rowther} S.,  {Nealon} R.,  {Meru} F.,  {Wurster} J.,  {Aly} H.,  {Alexander}
  R.,  {Rice} K.,   {Booth} R.~A.,  2024a, \mn@doi [\mnras]
  {10.1093/mnras/stae167}, \href
  {https://ui.adsabs.harvard.edu/abs/2024MNRAS.528.2490R} {528, 2490}

\bibitem[\protect\citeauthoryear{{Rowther}, {Price}, {Pinte}, {Nealon}, {Meru}
  \& {Alexander}}{{Rowther} et~al.}{2024b}]{rowther24b}
{Rowther} S.,  {Price} D.~J.,  {Pinte} C.,  {Nealon} R.,  {Meru} F.,
  {Alexander} R.,  2024b, \mn@doi [\mnras] {10.1093/mnras/stae2167}, \href
  {https://ui.adsabs.harvard.edu/abs/2024MNRAS.534.2277R} {534, 2277}

\bibitem[\protect\citeauthoryear{{Sch{\"a}fer}, {Yang}  \&
  {Johansen}}{{Sch{\"a}fer} et~al.}{2017}]{schafer17}
{Sch{\"a}fer} U.,  {Yang} C.-C.,   {Johansen} A.,  2017, \mn@doi [\aap]
  {10.1051/0004-6361/201629561}, \href
  {https://ui.adsabs.harvard.edu/abs/2017A&A...597A..69S} {597, A69}

\bibitem[\protect\citeauthoryear{{Schlaufman}}{{Schlaufman}}{2018}]{schlaufman18}
{Schlaufman} K.~C.,  2018, \mn@doi [\apj] {10.3847/1538-4357/aa961c}, \href
  {https://ui.adsabs.harvard.edu/abs/2018ApJ...853...37S} {853, 37}

\bibitem[\protect\citeauthoryear{{Shi}, {Zhu}, {Stone}  \& {Chiang}}{{Shi}
  et~al.}{2016}]{shi16}
{Shi} J.-M.,  {Zhu} Z.,  {Stone} J.~M.,   {Chiang} E.,  2016, \mn@doi [\mnras]
  {10.1093/mnras/stw692}, \href
  {https://ui.adsabs.harvard.edu/abs/2016MNRAS.459..982S} {459, 982}

\bibitem[\protect\citeauthoryear{{Simon}, {Armitage}, {Li}  \&
  {Youdin}}{{Simon} et~al.}{2016}]{simon16}
{Simon} J.~B.,  {Armitage} P.~J.,  {Li} R.,   {Youdin} A.~N.,  2016, \mn@doi
  [\apj] {10.3847/0004-637X/822/1/55}, \href
  {https://ui.adsabs.harvard.edu/abs/2016ApJ...822...55S} {822, 55}

\bibitem[\protect\citeauthoryear{{Steinacker} et~al.,}{{Steinacker}
  et~al.}{2015}]{steinacker2015}
{Steinacker} J.,  et~al., 2015, \mn@doi [\aap] {10.1051/0004-6361/201425434},
  \href {https://ui.adsabs.harvard.edu/abs/2015A&A...582A..70S} {582, A70}

\bibitem[\protect\citeauthoryear{{Toomre}}{{Toomre}}{1964}]{toomre64}
{Toomre} A.,  1964, \mn@doi [\apj] {10.1086/147861}, \href
  {https://ui.adsabs.harvard.edu/abs/1964ApJ...139.1217T} {139, 1217}

\bibitem[\protect\citeauthoryear{{Tu}, {Li}  \& {Lam}}{{Tu}
  et~al.}{2022}]{tu22}
{Tu} Y.,  {Li} Z.-Y.,   {Lam} K.~H.,  2022, \mn@doi [\mnras]
  {10.1093/mnras/stac2030}, \href
  {https://ui.adsabs.harvard.edu/abs/2022MNRAS.515.4780T} {515, 4780}

\bibitem[\protect\citeauthoryear{{Tychoniec} et~al.,}{{Tychoniec}
  et~al.}{2020}]{tychoniec20}
{Tychoniec} {\L}.,  et~al., 2020, \mn@doi [\aap] {10.1051/0004-6361/202037851},
  \href {https://ui.adsabs.harvard.edu/abs/2020A&A...640A..19T} {640, A19}

\bibitem[\protect\citeauthoryear{{Vigan} et~al.,}{{Vigan}
  et~al.}{2017}]{vigan17}
{Vigan} A.,  et~al., 2017, \mn@doi [\aap] {10.1051/0004-6361/201630133}, \href
  {https://ui.adsabs.harvard.edu/abs/2017A&A...603A...3V} {603, A3}

\bibitem[\protect\citeauthoryear{{Vorobyov}, {Kulikov}, {Elbakyan}, {McKevitt}
  \& {G{\"u}del}}{{Vorobyov} et~al.}{2024}]{vorobyov24}
{Vorobyov} E.~I.,  {Kulikov} I.,  {Elbakyan} V.~G.,  {McKevitt} J.,
  {G{\"u}del} M.,  2024, \mn@doi [\aap] {10.1051/0004-6361/202348023}, \href
  {https://ui.adsabs.harvard.edu/abs/2024A&A...683A.202V} {683, A202}

\bibitem[\protect\citeauthoryear{{Weidenschilling}}{{Weidenschilling}}{1977}]{weidenschilling77}
{Weidenschilling} S.~J.,  1977, \mn@doi [\mnras] {10.1093/mnras/180.2.57},
  \href {https://ui.adsabs.harvard.edu/abs/1977MNRAS.180...57W} {180, 57}

\bibitem[\protect\citeauthoryear{Yang \& Johansen}{Yang \&
  Johansen}{2016}]{yang16}
Yang C.-C.,  Johansen A.,  2016, \mn@doi [The Astrophysical Journal Supplement
  Series] {10.3847/0067-0049/224/2/39}, 224, 39

\bibitem[\protect\citeauthoryear{{Yang} \& {Krumholz}}{{Yang} \&
  {Krumholz}}{2012}]{yang12}
{Yang} C.-C.,  {Krumholz} M.,  2012, \mn@doi [\apj]
  {10.1088/0004-637X/758/1/48}, \href
  {https://ui.adsabs.harvard.edu/abs/2012ApJ...758...48Y} {758, 48}

\bibitem[\protect\citeauthoryear{{Youdin} \& {Johansen}}{{Youdin} \&
  {Johansen}}{2007}]{youdin07}
{Youdin} A.,  {Johansen} A.,  2007, \mn@doi [\apj] {10.1086/516729}, \href
  {https://ui.adsabs.harvard.edu/abs/2007ApJ...662..613Y} {662, 613}

\bibitem[\protect\citeauthoryear{{Youdin} \& {Lithwick}}{{Youdin} \&
  {Lithwick}}{2007}]{youdinlithwick07}
{Youdin} A.~N.,  {Lithwick} Y.,  2007, \mn@doi [\icarus]
  {10.1016/j.icarus.2007.07.012}, \href
  {https://ui.adsabs.harvard.edu/abs/2007Icar..192..588Y} {192, 588}

\makeatother
\end{thebibliography}








\bsp	
\label{lastpage}
\end{document}